\documentclass[draftclsnofoot, onecolumn, 11pt]{IEEEtran}
\ifCLASSINFOpdf
\else
\fi

\usepackage{amssymb}
\usepackage{latexsym}
\usepackage{graphicx}
\usepackage{color}
\usepackage{multirow}
\usepackage{amsmath}

\usepackage{amsfonts}

\newcommand{\abu}{{\bf a}}
\newcommand{\bbu}{{\bf b}}
\newcommand{\cbu}{{\bf c}}
\newcommand{\dbu}{{\bf d}}
\newcommand{\fbu}{{\bf f}}
\newcommand{\hbu}{{\bf h}}

\newcommand{\ubu}{{\bf u}}

\newcommand{\gbu}{{\bf g}}
\newcommand{\xbu}{{\bf x}}

\newcommand{\zbu}{{\bf z}}

\newcommand{\Cbu}{{\bf C}}

\newcommand{\Hbu}{{\bf H}}
\newcommand{\Ibu}{{\bf I}}
\newcommand{\Gbu}{{\bf G}}
\newcommand{\Tbu}{{\bf T}}

\newcommand{\qed}{\hfill \ensuremath{\Box}}

\newtheorem{df}{Definition}
\newtheorem{thr}{Theorem}
\newtheorem{lem}{Lemma}
\newtheorem{rem}{Remark}
\newtheorem{cor}{Corollary}
\newtheorem{const}{Construction}
\newtheorem{exa}{Example}

\begin{document}
\title{Reed-Muller Codes for Peak Power Control in Multicarrier CDMA}

%\author{\IEEEauthorblockN{Nam Yul Yu\IEEEauthorrefmark{1} and Guang Gong\IEEEauthorrefmark{2}} \\
%\IEEEauthorblockA{\IEEEauthorrefmark{1}Department of Electrical Engineering, Lakehead University \\
%Thunder Bay, ON, CANADA \\
%Email: nam.yu@lakeheadu.ca
%}
%\IEEEauthorblockA{\IEEEauthorrefmark{2}Department of Electrical and Computer Engineering, University of Waterloo\\
% Thunder Bay, ON, CANADA \\
%Email: ggong@calliope.uwaterloo.ca
%}
%}

\author{\IEEEauthorblockN{Nam Yul Yu} \\
\IEEEauthorblockA{Department of Electrical Engineering, Lakehead University \\
Thunder Bay, ON, CANADA \\
Email: nam.yu@lakeheadu.ca}
%\IEEEauthorblockA{\IEEEauthorrefmark{2}Department of Electrical and Computer Engineering, University of Waterloo\\
% Thunder Bay, ON, CANADA \\
%Email: ggong@calliope.uwaterloo.ca
}


%Telephone: (800) 555--1212, Fax: (888) 555--1212}


\maketitle

\begin{abstract}
%\boldmath
Reed-Muller codes are studied for peak power control in multicarrier code-division multiple access (MC-CDMA)
communication systems.
In a coded MC-CDMA system, the information data multiplexed from users
is encoded by a Reed-Muller subcode
and the codeword is fully-loaded to Walsh-Hadamard spreading sequences.
The polynomial representation of a coded MC-CDMA signal is established for
theoretical analysis of the peak-to-average power ratio (PAPR).
The Reed-Muller subcodes are defined in a recursive way by the Boolean functions
providing the transmitted MC-CDMA signals with the bounded PAPR
as well as the error correction capability.
A connection between the code rates and the maximum PAPR is theoretically investigated
in the coded MC-CDMA.
Simulation results present the statistical evidence that
the PAPR of the coded MC-CDMA signal is not only theoretically bounded, but also statistically reduced.
In particular, the coded MC-CDMA
solves the major PAPR problem of uncoded MC-CDMA
by dramatically reducing its PAPR
for the small number of users.
Finally, the theoretical and statistical studies show that
the Reed-Muller subcodes are effective coding schemes for peak power control in MC-CDMA
with small and moderate numbers of users, subcarriers, and spreading factors.
\end{abstract}


% \fi
%\IEEEpeerreviewmaketitle

\section{Introduction}
Multicarrier communications have recently attracted much attention % from researchers
in wireless and mobile applications.
The orthogonal frequency division multiplexing (OFDM) has been employed as a multiplexing and a multiple access technique
in a variety of wireless communication standards
such as IEEE802.11 wireless LAN~\cite{802.11}, IEEE802.16 mobile WiMAX~\cite{WiMAX},
and 3GPP-LTE~\cite{LTE:36-211}.
Also, the multicarrier code-division multiple access (MC-CDMA),
a combined scheme of OFDM and CDMA~\cite{Yee:MC-CDMA}$-$\cite{Hara:MC-CDMA},
has been proposed
to enjoy the benefits of OFDM and CDMA by allocating the spread data symbols to subcarriers.
The popularity of multicarrier communications is mainly due to the robustness to multipath fading channels
and the efficient hardware implementation employing fast Fourier transform (FFT) techniques.
However, multicarrier communications have the major drawback of
the high peak-to-average power ratio (PAPR) of transmitted signals,
which may nullify all the potential benefits~\cite{Pat:GRM}.

A number of techniques have been developed for
PAPR reduction of OFDM signals.
In particular, a constructive and theoretical approach is to employ a coding scheme~\cite{Jones:block}\cite{Nee:OFDM}
that provides low PAPR and good error correction capability for transmitted OFDM signals.
The Golay complementary sequences~\cite{Golay},
which belong to
a coset of the first-order Reed-Muller code~\cite{DavJed:GDJ},
are a good example of the coding scheme. % for OFDM signals.
Paterson~\cite{Pat:Codes} also discussed several coding schemes for PAPR reduction of \emph{multicode} CDMA.
In~\cite{Pat:alg}, he summarized the algebraic coding approaches for peak power control in OFDM and multicode CDMA.
For a summary of the other PAPR reduction techniques for OFDM, see~\cite{Han:overview}.

To reduce the PAPR of multicarrier CDMA (MC-CDMA) signals, on the other hand,
numerous studies have been focused on the power characteristics of spreading sequences.
Ochiai and Imai~\cite{OchImai:OFDM-CDMA} presented statistical results of the PAPR in downlink MC-CDMA,
where multiple users are supported by Walsh-Hadamard or Golay complementary spreading sequences.
Considering a single user MC-CDMA, Popovi\'c~\cite{Pop:spreading} presented the basic criteria
for the selection of spreading sequences
by studying the crest factors (CF) $- \ \sqrt{{\rm PAPR}} \ -$ of various binary and polyphase sequences.
Similar studies can be found in \cite{Nob:SS} with multiple access interference (MAI) minimization.
In MC-CDMA supporting multiple users or code channels, the crest factors of various spreading sequences have been compared
in \cite{ChoiHanzo:CF} and \cite{ChoiHanzo:CFCS},
where the Walsh-Hadamard spreading sequences showed the best PAPR properties, provided that a large number of
spreading sequences are combined for the transmitted MC-CDMA signals.
More studies can be found in
\cite{Pog:SS}$-$\cite{WieWu:High} on the PAPR of various spreading sequences in MC-CDMA.
%\cite{Pog:SS}, \cite{Yang:alloc}, \cite{Shi:sym}, and \cite{WieWu:High}.

If MC-CDMA assigns multiple spreading sequences to a single user,
the multicode MC-CDMA can be equivalently treated as the \emph{spread OFDM} (S-OFDM)~\cite{Kaiser:SOFDM}$-$\cite{Al:CSOFDM},
where a data symbol of the user is spread across a set of subcarriers % by spreading sequences
to enjoy frequency diversity. % by spreading the data symbols across a set of subcarriers.
%Furthermore, the Walsh-Hadamard transform (WHT) allows efficient spreading and despreading processes in Walsh-Hadamard spread OFDM.
If the number of used spreading sequences is large,
the Walsh-Hadamard spread OFDM can be viewed as a PAPR reducing scheme~\cite{ChoiHanzo:CF}\cite{Hanzo:book},
compared to a conventional OFDM.
Also, an error correction code may be applied prior to Walsh-Hadamard spreading
for improving the error rate performance~\cite{Al:CSOFDM}\cite{Ye:FEC} or controlling the peak power~\cite{ChoiHanzo:CF}
of S-OFDM.

To the best of our knowledge, most of the efforts on PAPR reduction of MC-CDMA and S-OFDM
have been verified mainly by statistical experiments, not by thorough theoretical analysis.
Through the experiments,
the PAPR of the multicarrier signals has been statistically observed, % through the efforts,,
but it has never been addressed whether it is theoretically bounded.
%Only a theoretical study can be found in~\cite{Parker:close},
%which made a general and mathematical study for PAPR, not considering the application to multicarrier CDMA.
In this paper, we propose a binary Reed-Muller coded MC-CDMA system and
study its PAPR properties. In the coded MC-CDMA,
the information data multiplexed from users is encoded by
a Reed-Muller subcode and the codeword is then fully-loaded to % all the available
Walsh-Hadamard spreading sequences.
The coding scheme plays a role of reducing the PAPR of transmitted MC-CDMA signals as well as
providing the error correction capability.
%First of all,
We first establish the polynomial representation of a coded MC-CDMA signal
for theoretical analysis of the PAPR.
A recursive construction of Boolean functions is then presented for the Reed-Muller subcodes, where
the PAPR of the MC-CDMA signal encoded by the subcode is proven to be theoretically bounded.
The author of \cite{Parker:close} pointed out that the construction is equivalent to
Type-III sequences in \cite{Parker:close} where he made a general and mathematical study for Boolean functions with bounded PAPR,
not considering the application to MC-CDMA.
We also discuss a connection between the code rate of the subcode and the maximum PAPR. % of the coded MC-CDMA signal.
Simulation results show that the PAPR of the coded MC-CDMA signal
is not only theoretically bounded, but also
statistically reduced. % by the Reed-Muller subcodes.
In particular, the coded MC-CDMA
solves the major PAPR problem of uncoded MC-CDMA
by dramatically reducing its PAPR % uncoded ones in theoretical and statistical aspects
for the small number of users. % supported by the system,
%the PAPR is dramatically reduced
%when the small number of users access to the coded MC-CDMA.
In conclusion, the Reed-Muller codes can be effectively utilized for peak power control in MC-CDMA
with small and moderate numbers of users, subcarriers, and spreading factors.
The PAPR properties of the coded MC-CDMA equivalently address those of the Reed-Muller coded S-OFDM % (or multicode MC-CDMA)
which supports multiple data from a single user.
%We believe this work gives theoretical insights
%for PAPR reduction of MC-CDMA and S-OFDM systems
%by means of an error correction coding.

%The rest of this paper is organized as follows.
%Section II describes a general coded MC-CDMA transmitter
%and introduces basic notations and concepts used in this paper.
%In Section III, we present the polynomial representation of a coded MC-CDMA signal,
%which allows the efficient PAPR analysis.
%In Section IV, we introduce main results of this paper by presenting
%a recursive construction of Reed-Muller subcodes
%and providing the theoretical PAPR bounds for the coded MC-CDMA signals.
%Also, the encoding and the decoding procedures are briefly discussed in Section IV.
%Section V shows the simulation results of the PAPR of coded MC-CDMA signals
%and compares them to those of uncoded MC-CDMA signals.
%Concluding remarks are given in Section VI.

\section{System Description}
Throughout this paper, MC-CDMA abbreviates \emph{multicarrier} CDMA $-$ not multicode CDMA.
This paper discusses a coded MC-CDMA system employing binary codewords, binary spreading sequences, and BPSK modulation.
Hence, we focus our description on binary cases.
The following notations will be used throughout this paper.

\begin{itemize}
\item[$-$]
$L$ is a spreading factor or spreading sequence length.
%Also, we assume $L$ \emph{orthogonal} spreading sequences of length $L$.
\item[$-$]
$w$ and $W$ are the actual and the maximum numbers of users
supported by a coded MC-CDMA system, respectively, where $w \leq W$.
In the rest of this paper, we will use the context of $w$ \emph{users},
where the $w$ users can be treated as $w$ data bits of a single user in S-OFDM or
multicode MC-CDMA.
\item[$-$]
%Each information bit from the $w$ users forms uncoded bits of length $W$, where zeros are appended if $w \leq W$.
%Then, the $W$-bit data is encoded by a $(K, W)$ code
%to produce a $K$-bit codeword, where $K \geq W$.
%Thus,
$K$ is the codeword length of a $(K, W)$ code, where $K \geq W$.
%each coded bit is spread by a spreading sequence of length $L$, where $K \leq L$.
\item[$-$]
$N$ is the number of information bits that
each user transmits in an OFDM symbol.
%Hence, a spreading process is accomplished $N$ times in an OFDM symbol for each user.
\item[$-$]
$\abu^{(i)} = (a_0 ^{(i)}, a_1 ^{(i)}, \cdots, a_{N-1} ^{(i)})$ denotes the
$N$-bit information of the $i$th user, $0 \leq i \leq w-1$,
while $\abu_n = (a_n ^{(0)}, a_n ^{(1)}, \cdots, a_n ^{(w-1)}, 0, \cdots, 0)$
denotes the $W$-bit uncoded data multiplexed from $w$ users and zero-tailed
at the $n$th spreading process, $0 \leq n \leq N-1$.
Note $a_n ^{(i)} \in \{ 0, 1 \}$.
\item[$-$]
%$\bbu^{(k)} = (b_0 ^{(k)}, b_1 ^{(k)}, \cdots, b_{N-1} ^{(k)})$ denotes
%a set of $N$-bit coded data of the $k$th user,
%while
$\bbu_n = (b_n ^{(0)}, b_n ^{(1)}, \cdots, b_n ^{(K-1)})$
denotes the coded output of $\abu_n$ by a $(K, W)$ code at the $n$th spreading process.
%for $0 \leq n \leq N-1$.
%a set of $K$-bit coded data from all users that experiences the $n$th spreading process.
Note $b_n ^{(k)} \in \{ 0, 1 \}$ for $0 \leq k \leq K-1$.
\item[$-$]
%$\dbu^{(k)} = (d_0 ^{(k)}, d_1 ^{(k)}, \cdots, d_{N-1} ^{(k)})$ denotes
%a set of $N$-bit BPSK-modulated data of the $k$th user,
%while
$\dbu_n = (d_n ^{(0)}, d_n ^{(1)}, \cdots, d_n ^{(K-1)})$
denotes the BPSK modulation output of $\bbu_n$ that experiences the $n$th spreading process. % for $0 \leq n \leq N-1$.
Hence, $d_n ^{(k)} = (-1)^{b_n ^{(k)} } \in \{ -1, +1 \}$. % for $0 \leq k \leq K-1$.
%a set of $K$-bit BPSK-modulated data from all users that experiences the $n$th spreading process.
%$\dbu^{(k)}$ and $\dbu_n$ are the BPSK-modulations of $\bbu^{(k)}$ and $\bbu_n$, respectively, so
\item[$-$]
$\cbu^{(k)} = (c_0 ^{(k)}, c_1 ^{(k)}, \cdots, c_{L-1} ^{(k)})$ denotes the
$L$-chip spreading sequence assigned for the $k$th coded bit of $\dbu_n$,
while $\cbu_l = (c_l ^{(0)}, c_l ^{(1)}, \cdots, c_l ^{(K-1)})^T$
is a set of the $l$th spreading chips across all $K$ spreading sequences, where $0 \leq l \leq L-1$.
%$0 \leq k \leq K-1$ and $0 \leq l \leq L-1$.
$\Cbu$ is a $K \times L$ orthogonal spreading matrix with $L \geq K$,
where $\cbu^{(k)}$ is the $k$th row vector
and $\cbu_l$ is the $l$th column vector.
\item[$-$]
$\ubu_n = (u_n ^{(0)}, u_n ^{(1)}, \cdots, u_n ^{(L-1)})$
denotes the output data of length $L$ of the $n$th spreading process.
%where
%each bit in $\dbu_n$ is spread and combined.
%a set of $K$-bit BPSK-modulated data from all users that experiences the $n$th spreading process.
%$\dbu^{(k)}$ and $\dbu_n$ are the BPSK-modulations of $\bbu^{(k)}$ and $\bbu_n$, respectively, so

\end{itemize}

\subsection{Coded MC-CDMA transmitter}

\begin{figure}
\centering
\input{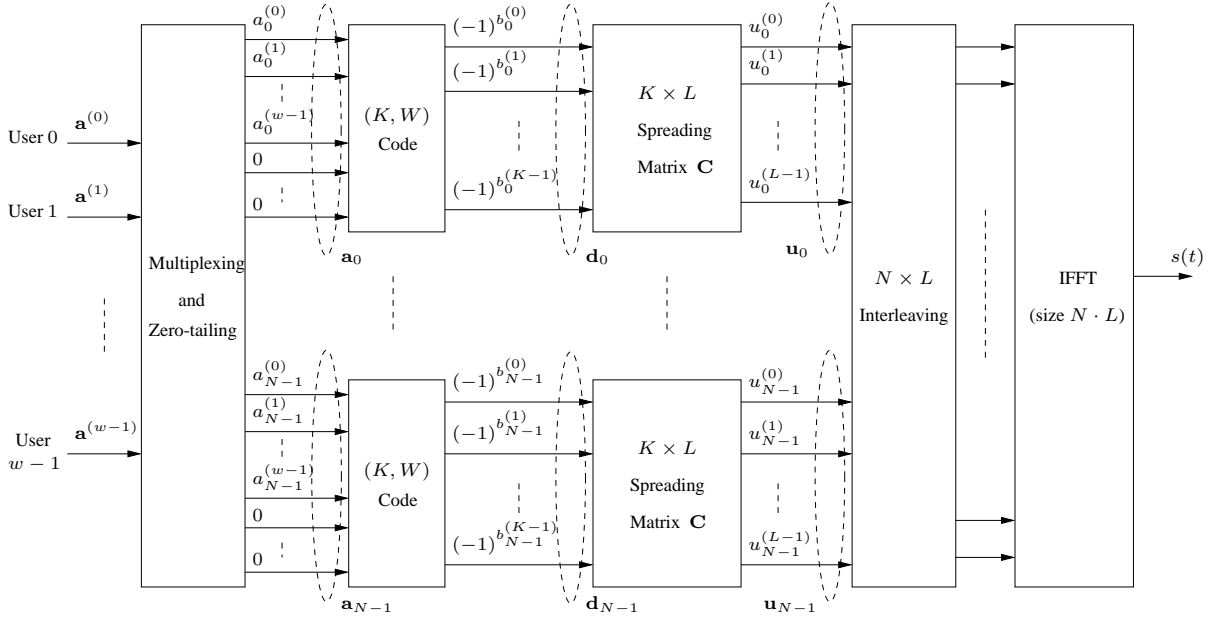}
\caption{A general coded MC-CDMA transmitter}
\label{fig:coded_sys}
\end{figure}

Figure~\ref{fig:coded_sys} illustrates a coded MC-CDMA transmitter proposed in this paper.
Assume that $w$ users access to the coded MC-CDMA system, where
the $i$th user, $0 \leq i \leq w-1$, is actively transmitting the $N$-bit information $\abu^{(i)}$ over an OFDM symbol.
The $n$th information bit of $\abu^{(i)}$ %from the $k$th user
is multiplexed across all $i$'s and then if $w \leq W$, zeros are attached to form the $W$-bit uncoded data $\abu_n$, $0 \leq n \leq N-1$.
Then, $\abu_n$ is encoded by a $(K, W)$ code to generate the $K$-bit codeword $\bbu_n$ and its BPSK modulation $\dbu_n$.
The $k$th coded bit of $\dbu_n$ is then spread by the $L$-bit spreading sequence $\cbu^{(k)}$, $0 \leq k \leq K-1$,
where a pair of the spreading sequences is mutually orthogonal.
At the $n$th spreading process, the spread bits of length $L$ are linearly combined over $K$ spreading sequences to produce $\ubu_n$,
where each element of $\ubu_n$ can take an arbitrary value.
Obviously, the spreading process is equivalent to a \emph{transform} of $\dbu_n$ by the orthogonal spreading matrix $\Cbu$, i.e.,
$\ubu_n = \dbu_n \cdot \Cbu$,
where $\cbu^{(k)}$ is the $k$th row vector of $\Cbu$.
%In particular, if $K=L$, then all the $L$ coded bits are \emph{fully-loaded} to
%all the available spreading sequences of length $L$,
%which converts the spreading process into a \emph{transform} by an orthogonal spreading matrix $\Cbu$.
The $N$ blocks of the spread data $\ubu_n$ of length $L$ experience an $N \times L$ block interleaver for frequency diversity,
and total $N \cdot L$ bits are allocated to $N \cdot L$ subcarriers by inverse FFT (IFFT).

The MC-CDMA receiver accomplishes the reverse operation to recover the original information $\abu^{(i)}$ for the $i$th user, where
%if it is a fully-loaded system, i.e., $K=L$,
the despreading process is equivalent to a transform by $\Cbu^{T}$, the transpose of $\Cbu$.

From Figure~\ref{fig:coded_sys}, the baseband transmission signal over an OFDM symbol duration $T_s$
is given by
\begin{equation}\label{eq:s_t}
s(t) = \sqrt{\frac{w}{K}} \cdot \sum_{n=0} ^{N-1} \sum_{l=0} ^{L-1} \sum_{k=0} ^{K-1} d_n ^{(k)} c_l ^{(k)} e^{j 2 \pi (Nl+ n)t/T_s },
\quad 0 \leq t < T_s
\end{equation}
where $j = \sqrt{-1}$. Note that if the zero-tail processes and the $(K, W)$ encoders are removed from Figure~\ref{fig:coded_sys},
then $s(t)$ with $K=w$ is equivalent to a conventional uncoded MC-CDMA signal in~\cite{OchImai:OFDM-CDMA}.
The normalization factor $\sqrt{\frac{w}{K}}$ is used in (\ref{eq:s_t}) to ensure that the average power of $s(t)$
is equal to that of the uncoded MC-CDMA signal for $w$ users, which will be shown in Section II-D.

%In this work, $\bbu_n$ is a codeword of length $2^m$ in a Reed-Muller subcode,
%and each codeword is \emph{fully-loaded} to all the available spreading sequences of length $2^m$.
%Therefore, $K=L=2^m$ in this paper.

\subsection{Boolean functions and Reed-Muller codes}
%As a $(K, W)$ code in Figure~\ref{fig:coded_sys}, we employ a subcode of the Reed-Muller code defined
%by a Boolean function.
Let $\xbu = (x_0, \cdots, x_{m-1})$ be a binary vector where $x_i \in \{0,1\}$, $0 \leq i \leq m-1$.
A \emph{Boolean function} $f(\xbu)$ is defined by
%a mapping $f: \Z_2 ^m \rightarrow \Z_q$,
%which is represented by the sum of all possible products of $x_{l}$'s with
%coefficients in $\Z_q$, i.e.,
\begin{equation}\label{eq:gBool}
f(\xbu) = f(x_0, \cdots, x_{m-1}) = \sum_{i=0} ^{2^m-1} v_i \prod_{l=0} ^{m-1} x_l ^{i_l}
\end{equation}
where $v_i \in \{0, 1\}$ and $i_l$ is obtained by a binary representation of $i = \sum_{l=0} ^{m-1} i_l 2^l$,
$i_l \in \{0, 1\}$.
Note that the addition in a Boolean function % in (\ref{eq:gBool})
is computed modulo-$2$.
%(\ref{eq:gBool}) is also called the \emph{algebraic normal form} of $f$.
In (\ref{eq:gBool}), the order of the $i$th monomial with nonzero $v_i$ is given by $\sum_{l=0} ^{m-1} i_l$,
and the highest order of the monomials with nonzero $v_i$'s is
called the \emph{(algebraic) degree} of the Boolean function $f$. % denoted by ${\rm deg}(f)$.

Associated with a Boolean function $f$, a binary codeword of length $2^m$
is defined by
\begin{equation}\label{eq:assoc_gen}
\fbu = (f_0, f_1, \cdots, f_{2^m-1}) \mbox{ where }
f_j = f(j_0 , j_1, \cdots, j_{m-1}), \ j= \sum_{l=0} ^{m-1} j_l 2^l
\end{equation}
where $j_l \in \{ 0, 1 \}$.
In other words, the \emph{associated} codeword $\fbu$ of length $2^m$ is obtained
by the Boolean function $f_j$
while $j$ runs through $0$ to $2^m-1$ in the increasing order.
%Note that a generalized Boolean function $f$ can generate all possible $q$-ary sequences of length $2^m$
%by all possible $c_i$'s in~(\ref{eq:gBool}).

The \emph{$r$th-order Reed-Muller code} ${\rm R}(r, m)$ is defined
by a set of binary codewords of length $2^m$ where each codeword is generated by a
Boolean function of degree at most $r$.
In other words,
each codeword in ${\rm R}(r, m)$ is the associated codeword $\fbu$ of length $2^m$ in (\ref{eq:assoc_gen})
where the Boolean function $f$ has the degree of at most $r$.
The $r$th-order Reed-Muller code ${\rm R}(r,m)$ has the dimension of $1+ \binom{m}{1} + \cdots + \binom{m}{r}$
and the minimum Hamming distance of $2^{m-r}$.

For more details on Boolean functions and Reed-Muller codes, see~\cite{Mac:ECC}.

\subsection{Walsh-Hadamard spreading sequences}
% In our coded MC-CDMA, we employ the Walsh-Hadamard matrix as a spreading matrix $\Cbu$.
The \emph{Walsh-Hadamard matrix} is recursively constructed by
$\Hbu_1 = [1]$ and
\begin{equation}\label{eq:WH}
\Hbu_{2^m} = \frac{1}{\sqrt{2}}
\begin{bmatrix}
\Hbu_{2^{m-1}} & \Hbu_{2^{m-1}} \\
\Hbu_{2^{m-1}} & -\Hbu_{2^{m-1}} \\
\end{bmatrix}.
\end{equation}
Then, it is easy to see that
\begin{equation}\label{eq:WH_prop}
\Hbu_{2^m} ^T = \Hbu_{2^m}, \quad
\Hbu_{2^m} \cdot \Hbu_{2^m} = \Ibu_{2^m}
\end{equation}
where $T$ denotes a transpose and $\Ibu_{2^m}$ the $2^m \times 2^m$ identity matrix.
(\ref{eq:WH_prop}) shows that the Walsh-Hadamard matrix is symmetric and orthogonal,
where the rows (or columns) are orthogonal vectors of length $2^m$, called
\emph{Walsh-Hadamard sequences}.
A theoretically defined Walsh-Hadamard matrix~\cite{SebYam:Had} has no normalization factor $\frac{1}{\sqrt{2}}$ in (\ref{eq:WH}).
However, we introduce it so that each Walsh-Hadamard sequence has the unit energy.

The Walsh-Hadamard sequences are described by the algebraic structure of
Boolean functions and the first-order Reed-Muller code.
In the Walsh-Hadamard matrix $\Hbu_{2^m}$, let $\hbu_l$ be the $l$th column vector of length $2^m$, i.e.,
$\hbu_l =
(h_{0,l}, h_{1,l}, \cdots , h_{2^m-1,l})^T $.
Let $l = \sum_{i=0} ^{m-1} l_i 2^i$ and $k = \sum_{i=0} ^{m-1} k_i 2^i$,
where $l_i , \ k_i \in \{ 0, 1 \}$.
Then, $h_{k,l}$ is given by
\begin{equation}\label{eq:h_col}
h_{k, l} = \frac{1}{\sqrt{2^m}} \cdot (-1)^{f_l (k_0, k_1, \cdots, k_{m-1})}
= \frac{1}{\sqrt{2^m}} \cdot (-1)^{\sum_{i=0} ^{m-1} l_i k_i}
\end{equation}
where the addition in the exponent is the modulo-$2$ addition.
Without the normalization factor $\frac{1}{\sqrt{2^m}}$,
the $l$th column vector of length $2^m$ is a `$\pm 1$'-codeword of length $2^m$
associated with the Boolean function $f_l$ of $m$ variables.
Precisely, $f_l$ generates $\hbu_l$ through (\ref{eq:h_col})
while the row index $k$ runs through $0$ to $2^m-1$,
where $\hbu_l$ is equivalent to a `$\pm 1$'-codeword in the first-order Reed-Muller code ${\rm R}(1, m)$.
%For more details on Boolean functions and Reed-Muller codes, see~\cite{Mac:ECC}.
Since each column vector corresponds to a codeword of ${\rm R}(1, m)$ of the minimum Hamming weight $2^{m-1}$,
it is straightforward that the sum of the column elements is either $\sqrt{2^m}$ or $0$, i.e.,
\begin{equation}\label{eq:exp_sum}
\sum_{k=0} ^{2^m-1} h_{k, l} = \frac{1}{\sqrt{2^m}}  \sum_{k=0} ^{2^m-1} (-1)^{\sum_{i=0} ^{m-1} l_i k_i}
= \left\{ \begin{array}{ll} \sqrt{2^m}, & \quad \mbox{if } l=0 \\
0, & \quad \mbox{if } l \neq 0 \end{array} \right.
\end{equation}

\vspace{0.1in}
\begin{exa}
From (\ref{eq:WH}), a $4 \times 4$ Walsh-Hadamard matrix is given by
\begin{equation*}
\Hbu_4 = \frac{1}{\sqrt{4}}
\begin{bmatrix}
1 & 1 & 1 & 1 \\
1 & -1 & 1 & -1 \\
1 & 1 & -1 & -1 \\
1 & -1 & -1 & 1 \\
\end{bmatrix}
=
\begin{bmatrix}
\hbu_{0} & \hbu_{1} & \hbu_{2} & \hbu_{3} \\
\end{bmatrix} .
%& = \frac{1}{\sqrt{4}}
%\begin{bmatrix}
%(-1)^0 & (-1)^{k_0} & (-1)^{k_1} & (-1)^{k_0+k_1} \\
%\end{bmatrix}
\end{equation*}
%For the $l$th column vector $\hbu_l$, $0 \leq l \leq 3$,
We have $f_0(k_0, k_1) = 0$, $f_1(k_0, k_1) = k_0$, $f_2(k_0, k_1) = k_1$, and
$f_3(k_0, k_1) = k_0+k_1$.
Then, it is easily checked that
each Boolean function $f_l$ generates the $l$th column vector $\hbu_l$, $0 \leq l \leq 3$, through (\ref{eq:h_col})
while the row index $k = \sum_{i=0} ^{1} k_i 2^i$ runs through $0 $ to $3$.
Also, (\ref{eq:exp_sum}) is true for each column vector.
\end{exa}
\vspace{0.1in}

The \emph{Walsh-Hadamard transform}~\cite{Mac:ECC} of a vector $\gbu = (g_0, \cdots, g_{2^m-1})$ is
defined by
\[
\widehat{g}_l = \frac{1}{\sqrt{2^m}} \sum_{k=0} ^{2^m-1} g_k \cdot (-1)^{\sum_{i=0} ^{m-1} l_i k_i}, \quad 0 \leq l \leq 2^m-1
\]
where $l = \sum_{i=0} ^{m-1} l_i 2^i$ and $k = \sum_{i=0} ^{m-1} k_i 2^i$.
%Since each column vector of the Walsh-Hadamard matrix is associated with a linear Boolean function,
From the algebraic structure of the Walsh-Hadamard matrix described above,
it is straightforward that the Walsh-Hadamard transform of $\gbu$ is given as % a matrix form by
\begin{equation*}\label{eq:WHT}
\widehat{\gbu} = (\widehat{g}_0, \cdots, \widehat{g}_{2^m-1}) = \gbu \cdot \Hbu_{2^m}.
\end{equation*}

\subsection{Peak-to-Average Power Ratio (PAPR)}
The peak-to-average power ratio (PAPR)~\cite{Lit:ppc} of a multicarrier signal $s(t)$ is defined by
\begin{equation*}\label{eq:PAPR}
{\rm PAPR} \left(s(t)\right) = \frac{\max_{0 \leq t < T_s} | s(t)|^2}{E[|s(t)|^2]}
\end{equation*}
where $T_s$ is an OFDM symbol duration and $E[\cdot]$ denotes the ensemble average.
Using the orthogonality of spreading sequences,
the approach made in~\cite{Hanzo:book} implies that
the average power of the MC-CDMA signal $s(t)$ in (\ref{eq:s_t}) is determined by
\begin{equation}\label{eq:avg_pwr}
E[|s(t)|^2] = \frac{w}{K} \cdot \sum_{n=0} ^{N-1} \sum_{l=0} ^{L-1} \sum_{k=0} ^{K-1} |d_n ^{(k)}|^2 |c_l ^{(k)}|^2 .
\end{equation}
In particular, if $d_n ^{(k)} \in \{ -1, +1 \}$ and $\cbu^{(k)}$ has the unit energy, i.e., $\sum_{l=0} ^{L-1} |c_l ^{(k)}|^2 = 1$,
then
(\ref{eq:avg_pwr}) becomes
\begin{equation}\label{eq:avg_pwr2}
E[|s(t)|^2] = \frac{w}{K} \cdot N \cdot  K = Nw
\end{equation}
which is equal to
the average power of an uncoded MC-CDMA signal
where each of $w$ users transmits the $N$-bit information over an OFDM symbol duration. %~\cite{Hanzo:book}.

In the following, we define a polynomial $S(z)$ \emph{associated with} $s(t)$, similar to \cite{ParkKen:GCS}.

\vspace{0.1in}
\begin{df}\label{def:assoc_poly}
In general, the coded MC-CDMA signal $s(t)$ in (\ref{eq:s_t}) has a form of
$s(t) = \sum_{i=0} ^{NL-1} s_i e^{j 2 \pi i t/T_s } $, where
$N \cdot L$ is the number of subcarriers over an OFDM symbol and $s_i$ takes an arbitrary value.
With $z = e^{j 2 \pi t/T_s }$, the \emph{associated} polynomial $S(z)$ is defined by
\begin{equation}\label{eq:S_z_org}
S(z) = \sum_{i=0} ^{NL-1} s_i z^{i}.  %= \ubu \cdot \zbu = \dbu_0 \cdot \Cbu \cdot \zbu
\end{equation}
%From (\ref{eq:avg_pwr2}), $E[|S(z)|^2] = Nw$ for $z=e^{j 2 \pi t/T_s }$.
%where $\ubu = (u_0, \cdots, u_{L-1})$ and $\zbu = (1, z, z^2, \cdots, z^{L-1})^T$.
From (\ref{eq:avg_pwr2}) and (\ref{eq:S_z_org}), the PAPR of $s(t)$ is translated into
\begin{equation}\label{eq:S_z_papr}
{\rm PAPR}(s(t)) = \frac{\max_{|z|=1} |S(z)|^2}{Nw}.
\end{equation}
\end{df}
\vspace{0.1in}

\section{Polynomial Representation of A Coded MC-CDMA Signal}
%Throughout this paper,
%we assume a \emph{fully-loaded} MC-CDMA system,
%where all the available $L$ spreading sequences are assigned
%for $L$ coded data bits, so $K=L$.
We establish the polynomial representation of a coded MC-CDMA signal
by presenting the associated polynomial introduced in Definition~\ref{def:assoc_poly}.
For simplicity, we first study the polynomial representation
for $N=1$,
where each user transmits a single information bit
with a single spreading process in an OFDM symbol.
The general representation with $N>1$ is then discussed.
% by the interleaving
% of the signals for $N=1$.

\subsection{$N=1$}
%Theorem~\ref{th:matrix_s} describes a polynomial representation of a coded MC-CDMA signal for $N=1$.

With % $K=L$ and
$N=1$,
a coded MC-CDMA signal is denoted by
\begin{equation}\label{eq:s_t_0}
s_0(t) = \sqrt{\frac{w}{K}} \cdot  \sum_{l=0} ^{L-1} \sum_{k=0} ^{K-1} d_0 ^{(k)} c_l ^{(k)} e^{j 2 \pi l t/T_s },
\quad 0 \leq t < T_s.
\end{equation}
Then, the polynomial representation of $s_0(t)$ is established by the following theorem.

\vspace{0.1in}
\begin{thr}\label{th:matrix_s}
The polynomial $S_0(z)$ associated with $s_0(t)$ in (\ref{eq:s_t_0}) is given by
\begin{equation}\label{eq:S_z}
S_0(z) = \sqrt{\frac{w}{K}} \cdot \dbu_0 \cdot \Cbu \cdot \zbu
\end{equation}
where $\zbu = (1, z, z^2, \cdots, z^{L-1})^T$. % and $\Cbu$ is a $K \times L$ spreading matrix.
In particular, if $K=L=2^m$ and $\Cbu$ is a
$2^m \times 2^m$ Walsh-Hadamard matrix, i.e., $\Cbu = \Hbu_{2^m}$, then
%the polynomial $S_0(z)$ is
\begin{equation}\label{eq:S_z_Had}
S_0(z) = \sqrt{\frac{w}{2^m}} \cdot \dbu_0 \cdot \Hbu_{2^m} \cdot \zbu = \sqrt{\frac{w}{2^m}} \cdot \widehat{\dbu}_0 \cdot \zbu
% = \sqrt{\frac{w}{K}} \cdot \dbu_0 \cdot \Tbu
\end{equation}
where $\widehat{\dbu}_0 $ is the Walsh-Hadamard transform of $\dbu_0$. % in (\ref{eq:WHT}),
%and $\Tbu$ is given in (\ref{eq:Cz}).
\end{thr}
\vspace{0.1in}

\noindent \emph{Proof.}
%With $K=L$,
%$\dbu_0 = (d_0 ^{(0)}, \cdots, d_0 ^{(K-1)})$
%is a binary codeword of length $K$ where $d_0 ^{(k)} \in \{ -1, +1\}$, $0 \leq k \leq K-1$.
In (\ref{eq:s_t_0}), let
$s_0(t) = \sum_{l=0} ^{L-1} u_0 ^{(l)} e^{j 2 \pi l t/T_s } $
where $u_0 ^{(l)} = \sqrt{\frac{w}{K}} \cdot  \sum_{k=0} ^{K-1} d_0 ^{(k)} c_l ^{(k)}$.
Then,
\begin{equation}\label{eq:u_0}
\ubu_0 =  (u_0 ^{(0)}, \cdots, u_0 ^{(L-1)} ) = \sqrt{\frac{w}{K}} \cdot \dbu_0 \cdot \Cbu.
\end{equation}
With $z = e^{j 2 \pi t/T_s }$,
the associated polynomial is then given by
$ S_0(z) = \sum_{l=0} ^{L-1} u_0 ^{(l)} z^{l}  = \ubu_0 \cdot \zbu$, % = \sqrt{\frac{w}{K}} \cdot \dbu_0 \cdot \Cbu \cdot \zbu
which derives (\ref{eq:S_z}).
%where $\zbu = (1, z, z^2, \cdots, z^{L-1})^T$.
If $\Cbu = \Hbu_{2^m}$, then (\ref{eq:S_z_Had}) is immediate.
\qed

\vspace{0.1in}
\begin{cor}\label{co:S_z_sqr}
With % $K=L$ and
$N=1$, if
a coded MC-CDMA signal $s_0(t)$ has the PAPR of at most $P$, then
\[
\left| S_0(z) \right|^2 \leq w P
\]
from (\ref{eq:S_z_papr}).
Also, we have from (\ref{eq:S_z})
\begin{equation*}\label{eq:dCz}
\left| \dbu_0 \cdot \Cbu \cdot \zbu \right|^2 = \frac{K}{w} \cdot \left| S_0(z) \right|^2 \leq KP.
\end{equation*}
where $\zbu = (1, z, z^2, \cdots, z^{L-1})^T$ with $|z|=1$.
\end{cor}
\vspace{0.1in}

In (\ref{eq:u_0}), if $\Cbu = \Hbu_{2^m}$, the spread output
$\ubu_0$ is the Walsh-Hadamard transform of $\dbu_0$
with a scaling factor $\sqrt{\frac{w}{K}} $ in a Walsh-Hadamard spread MC-CDMA,
where $K=L=2^m$.
A similar property has been noticed
in a spreading process of multicode CDMA~\cite{Pat:Codes}.
%which turns out to be also useful for analyzing the PAPR of MC-CDMA signals.
In addition, $\Hbu_{2^m} \cdot \zbu$ % in $S_0(z)$
has the following structure.

%From the notations of (\ref{eq:inf_matrix}) and (\ref{eq:ss_matrix}),

%The associated polynomial $S(z)$ and the matrix representation (\ref{eq:S_z})
%are then used to analyze the PAPR of $s(t)$.

\vspace{0.1in}
\begin{lem}\label{lemma:Cz}
Let $\Hbu_{2^m}$ be a $2^m \times 2^m$ Walsh-Hadamard matrix % i.e., $\Cbu = \Hbu_{2^m}$,
and $\zbu = (1, z, z^2, \cdots, z^{2^m-1})^T$.
Then,
\begin{equation}\label{eq:Cz}
\Tbu % = \Cbu \cdot \zbu
= \Hbu_{2^m} \cdot \zbu =
\begin{bmatrix}
\phi_0 \phi_1 \cdots \phi_{m-2} \phi_{m-1} \\
\theta_0 \phi_1 \cdots \phi_{m-2} \phi_{m-1} \\
\phi_0 \theta_1 \cdots \phi_{m-2} \phi_{m-1} \\
\vdots \\
\theta_0 \theta_1 \cdots \theta_{m-2} \phi_{m-1} \\
\phi_0 \phi_1 \cdots \phi_{m-2} \theta_{m-1} \\
\theta_0 \phi_1 \cdots \phi_{m-2} \theta_{m-1} \\
\phi_0 \theta_1 \cdots \phi_{m-2} \theta_{m-1} \\
\vdots \\
\theta_0 \theta_1 \cdots \theta_{m-2} \theta_{m-1} \\
\end{bmatrix}
=
\begin{bmatrix}
G_0 (\phi_0, \cdots, \phi_{m-1}, \theta_0, \cdots, \theta_{m-1}) \\
G_1 (\phi_0, \cdots, \phi_{m-1}, \theta_0, \cdots, \theta_{m-1})\\
\vdots \\
G_{2^m-1} (\phi_0, \cdots, \phi_{m-1}, \theta_0, \cdots, \theta_{m-1}) \\
\end{bmatrix}
\end{equation}
where $G_i (\phi_0, \cdots, \phi_{m-1}, \theta_0, \cdots, \theta_{m-1}) =
\prod_{t=0} ^{m-1} \phi_t ^{\overline{i_t}} \theta_t ^{i_t}, \, i = \sum_{t=0} ^{m-1} i_t 2^t$,
where $\phi_t = \frac{1+z^{2^t}}{\sqrt{2}}$ and $\theta_t = \frac{1-z^{2^t}}{\sqrt{2}}$.
Note that $\overline{i_t} = 0$ if $i_t = 1$, or $\overline{i_t} = 1$ if $i_t = 0$.
\end{lem}
\vspace{0.1in}

\noindent \emph{Proof.}
If $m=1$, then
\begin{equation}\label{eq:m1}
\Hbu_{2} \cdot \zbu =
\frac{1}{\sqrt{2}}
\begin{bmatrix}
1 & 1 \\
1 & -1 \\
\end{bmatrix}
\begin{bmatrix}
1 \\
z \\
\end{bmatrix}
=
\begin{bmatrix}
\phi_0 \\
\theta_0 \\
\end{bmatrix}.
\end{equation}
Thus, (\ref{eq:Cz}) is true for $m=1$.
Assume (\ref{eq:Cz}) also holds for $m=k-1$, i.e.,
\begin{equation}\label{eq:m_k-1}
\Hbu_{2^{k-1}} \cdot \zbu_1 =
%\begin{bmatrix}
%\phi_0 \phi_1 \cdots \phi_{k-2}  \\
%\theta_0 \phi_1 \cdots \phi_{k-2}  \\
%%\phi_0 \theta_1 \cdots \phi_{k-2}  \\
%\vdots \\
%\theta_0 \theta_1 \cdots \phi_{k-2}  \\
%\phi_0 \phi_1 \cdots \theta_{k-2} \\
%\phi_0 \theta_1 \cdots \theta_{k-2} \\
%\vdots \\
%\theta_0 \theta_1 \cdots \theta_{k-2}  \\
%\end{bmatrix}
%=
\begin{bmatrix}
G_0 (\phi_0, \cdots, \phi_{k-2}, \theta_0, \cdots, \theta_{k-2}) \\
G_1 (\phi_0, \cdots, \phi_{k-2}, \theta_0, \cdots, \theta_{k-2})\\
\vdots \\
G_{2^{k-1}-1} (\phi_0, \cdots, \phi_{k-2}, \theta_0, \cdots, \theta_{k-2}) \\
\end{bmatrix}
\end{equation}
where $\zbu_1 = (1, z, z^2, \cdots, z^{2^{k-1}-1})^T$.
From the recursive construction of $\Hbu_{2^k}$,
we have
\begin{equation}\label{eq:m_k}
\Hbu_{2^k} \cdot \zbu = \frac{1}{\sqrt{2}}
\begin{bmatrix}
\Hbu_{2^{k-1}} & \Hbu_{2^{k-1}} \\
\Hbu_{2^{k-1}} & -\Hbu_{2^{k-1}} \\
\end{bmatrix}
\begin{bmatrix}
\zbu_1 \\
z^{2^{k-1}} \cdot \zbu_1 \\
\end{bmatrix}
=
\begin{bmatrix}
\left( \frac{1+z^{2^{k-1}}}{\sqrt{2}} \right) \cdot \Hbu_{2^{k-1}} \cdot \zbu_1 \\
\left( \frac{1-z^{2^{k-1}}}{\sqrt{2}} \right)  \cdot \Hbu_{2^{k-1}} \cdot \zbu_1  \\
\end{bmatrix}
=
\begin{bmatrix}
\phi_{k-1}  \cdot \Hbu_{2^{k-1}} \cdot \zbu_1 \\
\theta_{k-1} \cdot \Hbu_{2^{k-1}} \cdot \zbu_1  \\
\end{bmatrix}
\end{equation}
where $\zbu = (1, z, z^2, \cdots, z^{2^k-1})^T = (\zbu_1 ^T, z^{2^{k-1}}\cdot\zbu_1 ^T)^T$.
Thus, from (\ref{eq:m_k-1}) and (\ref{eq:m_k}), % it is immediate that
\begin{equation}\label{eq:m_k_true}
\Hbu_{2^k} \cdot \zbu =
%\begin{bmatrix}
%\phi_0 \phi_1 \cdots \phi_{k-1}  \\
%\theta_0 \phi_1 \cdots \phi_{k-1}  \\
%%\phi_0 \theta_1 \cdots \phi_{k-2}  \\
%\vdots \\
%\theta_0 \theta_1 \cdots \phi_{k-1}  \\
%\phi_0 \phi_1 \cdots \theta_{k-1} \\
%\phi_0 \theta_1 \cdots \theta_{k-1} \\
%\vdots \\
%\theta_0 \theta_1 \cdots \theta_{k-1}  \\
%\end{bmatrix}
\begin{bmatrix}
\phi_{k-1} \cdot G_0 (\phi_0, \cdots, \phi_{k-2}, \theta_0, \cdots, \theta_{k-2})  \\
\phi_{k-1} \cdot G_1 (\phi_0, \cdots, \phi_{k-2}, \theta_0, \cdots, \theta_{k-2})  \\
\vdots \\
\phi_{k-1} \cdot G_{2^{k-1}-1} (\phi_0, \cdots, \phi_{k-2}, \theta_0, \cdots, \theta_{k-2}) \\
\theta_{k-1} \cdot G_{0} (\phi_0, \cdots, \phi_{k-2}, \theta_0, \cdots, \theta_{k-2}) \\
\theta_{k-1} \cdot G_{1} (\phi_0, \cdots, \phi_{k-2}, \theta_0, \cdots, \theta_{k-2}) \\
\vdots \\
\theta_{k-1} \cdot G_{2^{k-1}-1} (\phi_0, \cdots, \phi_{k-2}, \theta_0, \cdots, \theta_{k-2}) \\
\end{bmatrix}
=
\begin{bmatrix}
G_0 (\phi_0, \cdots, \phi_{k-1}, \theta_0, \cdots, \theta_{k-1}) \\
G_1 (\phi_0, \cdots, \phi_{k-1}, \theta_0, \cdots, \theta_{k-1})\\
\vdots \\
G_{2^{k}-1} (\phi_0, \cdots, \phi_{k-1}, \theta_0, \cdots, \theta_{k-1}) \\
\end{bmatrix}.
\end{equation}
From (\ref{eq:m1}) and (\ref{eq:m_k_true}),
(\ref{eq:Cz}) is true by induction.
\qed

\vspace{0.1in}
\begin{rem}
In Definition 6 of~\cite{ParkTell:GDJ}, Parker and Tellambura defined
$\Gbu_m$, a set of normalized complex sequences
of length $2^m$ by a tensor product, where
$|\phi_t|^2  + |\theta_t|^2 = 2$.
In fact, $\Hbu_{2^m} \cdot \zbu$ in Lemma~\ref{lemma:Cz} is a special case of $\Gbu_m$
with $\phi_t = \frac{1+z^{2^t}}{\sqrt{2}}$ and $\theta_t = \frac{1-z^{2^t}}{\sqrt{2}}$,
$0 \leq t \leq m-1$.
\end{rem}
\vspace{0.1in}

\subsection{$N>1$}
%For $N>1$, the associated polynomial of
%a coded MC-CDMA signal $s(t)$ is determined in the following theorem.
In (\ref{eq:s_t_0}),
replacing $d_0 ^{(k)}$ by $d_n ^{(k)}$
leads us to $s_n(t)$ and its associated polynomial $S_n(z)$, i.e.,
%if $d_n ^{(k)}$ replaces $d_0 ^{(k)}$ from $s_0(t)$, % in (\ref{eq:s_t_0}),
%then $s_n(t)$ and its associated polynomial $S_n(z)$ are given by
\begin{equation}\label{eq:s_n_2}
s_n (t) = \sqrt{\frac{w}{K}} \cdot \sum_{l=0} ^{L-1} \sum_{k=0} ^{K-1} d_n ^{(k)} c_l ^{(k)} e^{j 2 \pi lt/T_s },
\quad S_n (z) = \sqrt{\frac{w}{K}} \cdot \dbu_n \cdot \Cbu \cdot \zbu,
\quad 0 \leq n \leq N-1
\end{equation}
where $\zbu = (1, z, z^2, \cdots, z^{L-1})^T$.
Obviously, $s_n(t)$ is also a coded MC-CDMA signal % in Figure~\ref{fig:coded_sys},
where $\dbu_n$ is loaded with a single spreading process over an OFDM symbol.
In particular, if $\Cbu=\Hbu_{2^m}$, then the spreading process is equivalent to the Walsh-Hadamard transform (WHT),
which enables the efficient implementation of spreading and despreading processes.
%Using $S_n(z)$, we determine the associated polynomial of a coded MC-CDMA signal in (\ref{eq:s_t}).

\vspace{0.1in}
\begin{thr}\label{th:S_z_N}
%Let $\widetilde{S}_n(z)$ be the associated polynomial of $\widetilde{s}_n(t)$ in (\ref{eq:s_n}), where
%$0 \leq n \leq N-1$.
%With $K=L$,
With $S_n(z)$ in (\ref{eq:s_n_2}), % be the associated polynomial of $s_n(t)$ in (\ref{eq:s_n_2}).
% Then,
the associated polynomial of a coded MC-CDMA signal $s(t)$ in (\ref{eq:s_t}) is determined by
\begin{equation}\label{eq:S_z_N}
S(z) = \sum_{n=0} ^{N-1} S_n (z^N) \cdot z^n .
\end{equation}
In other words, $S(z)$ is a polynomial obtained by
\emph{interleaving} $S_n(z)$'s for $0 \leq n \leq N-1$.
\end{thr}
\vspace{0.1in}

\noindent \emph{Proof.}
In (\ref{eq:s_t}),
\begin{equation}\label{eq:s_n}
\begin{split}
& s(t) = \sqrt{\frac{w}{K}} \cdot \sum_{n=0} ^{N-1} \sum_{l=0} ^{L-1} \sum_{k=0} ^{K-1} d_n ^{(k)} c_l ^{(k)} e^{j 2 \pi (Nl+ n)t/T_s }
= \sum_{n=0} ^{N-1} \widetilde{s}_n (t) \\
& \mbox{where }
\widetilde{s}_n (t) = \sqrt{\frac{w}{K}} \cdot \sum_{l=0} ^{L-1} \sum_{k=0} ^{K-1} d_n ^{(k)} c_l ^{(k)} e^{j 2 \pi (Nl+ n)t/T_s }.
\end{split}
\end{equation}
For a given $n$, $\widetilde{s}_n (t)$ is a signal assigned to the $(Nl+n)$th subcarriers
while $l$ runs through $0$ to $ L-1$.
From (\ref{eq:s_n}),
it is straightforward that the associated polynomial of $\widetilde{s}_n(t)$ is given by
\[
\widetilde{S}_n (z)  =  \sqrt{\frac{w}{K}} \cdot \dbu_n \cdot \Cbu \cdot \zbu^{(N)} \cdot z^n
\]
where $\zbu^{(N)} = (1, z^N, z^{2N}, \cdots, z^{(L-1)N})^T$.
Compared to (\ref{eq:s_n_2}),
%from $S_n (z) = \sqrt{\frac{w}{K}} \cdot \dbu_n \cdot \Cbu \cdot \zbu$,
%we have
$\widetilde{S}_n (z)  =  S_n(z^N) \cdot z^n$, and thus %from (\ref{eq:s_n_2}).
%Therefore,
the associated polynomial $S(z)$ is % given by
\[
S(z) = \sum_{n=0} ^{N-1} \widetilde{S}_n (z) =  \sum_{n=0} ^{N-1} S_n (z^N) \cdot z^n .
\]
\qed
\vspace{0.1in}

\begin{rem}\label{rm:cor1}
Since $s_n(t)$ is
a coded MC-CDMA signal with a single spreading process over an OFDM symbol,
Corollary~\ref{co:S_z_sqr} is also valid for $S_n(z)$. % in (\ref{eq:s_n_2}).
Precisely, if $s_n(t)$ has the PAPR of at most $P$, then % it implies that
$\left| S_n (z) \right|^2 \leq w P$ and $\left| \dbu_n \cdot \Cbu \cdot \zbu \right|^2 \leq KP$
for $0 \leq n \leq N-1$,
where $\zbu = (1, z, z^2, \cdots, z^{L-1})^T$ with $|z|=1$.
\end{rem}
\vspace{0.1in}

%Recall a coded MC-CDMA signal $s(t)$ in (\ref{eq:s_n}), where $K=L$ and $N > 1$.
%for a fully-loaded and Walsh-Hadamard spread MC-CDMA system with $K=L$ and $N \geq 1$.
Using its associated polynomial $S(z)$ in Theorem~\ref{th:S_z_N},
we determine the PAPR bound of a coded MC-CDMA signal $s(t)$ with $N>1$.

\vspace{0.1in}
\begin{thr}\label{th:papr_s_n}
%In a fully-loaded MC-CDMA with $K=L$,
In (\ref{eq:s_n_2}),
assume the maximum PAPR of $s_n(t)$ is $P$, i.e.,
$\max_{0 \leq n \leq N-1} {\rm PAPR}(s_n(t)) = P$.
Then, the coded MC-CDMA signal $s(t)$ in~(\ref{eq:s_t}) has the PAPR of at most $NP$, i.e.,
\[
{\rm PAPR}(s(t)) \leq  NP.
\]
\end{thr}

\noindent \emph{Proof.}
From the associated polynomial $S(z)$ in (\ref{eq:S_z_N}),
\begin{equation*}
\begin{split}
|S(z)|^2 & = \left| S_0 (z^N) + S_1 (z^N) \cdot z + \cdots + S_{N-1} (z^N) \cdot z^{N-1} \right|^2  \\
& \leq \sum_{n=0} ^{N-1} |z^n|^2 \cdot \sum_{n=0} ^{N-1} \left|S_n (z^N) \right|^2 \\
%& \leq \left| 1+ z+ \cdots + z^{N-1} \right|^2 \cdot \max_{0 \leq n \leq N-1} \left|S_n (z^N) \right|^2 \\
& \leq N^2 \cdot \max_{0 \leq n \leq N-1} \left|S_n (z^N) \right|^2
\end{split}
\end{equation*}
where $|z|=1$.
From Remark~\ref{rm:cor1},
$\max_{0 \leq n \leq N-1} {\rm PAPR}(s_n(t)) = P$ implies $\left| S_n (z^N) \right|^2 \leq wP$
for every $n$. Thus, % we have
$ \left|S (z) \right|^2 \leq N^2 \cdot w P$.
Therefore, % from (\ref{eq:S_z_papr}),
the PAPR of $s(t)$ is bounded by
\[
{\rm PAPR}( s(t))  = \frac{\max_{|z|=1} \left| S(z) \right|^2}{ N  w}
\leq \frac{N^2 \cdot w P}{N  w} = N P.
\]
\qed
\vspace{0.1in}

Although the proof is straightforward and the bound seems not so tight,
Theorem~\ref{th:papr_s_n} gives us an insight that
the maximum PAPR of coded MC-CDMA signals increases
as each user transmits more data bits ($N$) in an OFDM symbol.
Therefore, $N$ should be as small as possible
to remove the probability that the MC-CDMA signal has the high PAPR. % close to the theoretical maximum.
%In fact, statistical results will be provided in next section,
%where if $w$ is large, most of the MC-CDMA signals have much smaller PAPR than the maximum predicted in Theorem~\ref{th:papr_s_n}.

\section{Reed-Muller Codes for MC-CDMA}
In this section,
we develop a variety of subcodes of ${\rm R}(r, m)$ for a $(K, W)$ coding scheme of a coded MC-CDMA in
Figure~\ref{fig:coded_sys}, where
the codeword of length $K = 2^m$ is associated with a Boolean function of degree $r$.
We assume that the $K$-bit codeword is \emph{fully-loaded} to
all the available Walsh-Hadamard spreading sequences of length $2^m$, so $K=L=2^m$.
We analyze the PAPR properties of the fully-loaded, Reed-Muller coded, and Walsh-Hadamard spread MC-CDMA signals.
First of all, we study the PAPR for $N=1$. % for simplicity.
Then, the PAPR for $N > 1$ is investigated. % by repeatedly applying the code for the $n$th spreading process.

In the MC-CDMA system with $N=1$,
%$\dbu_0$ is a BPSK-modulated codeword of length $2^m$,
%encoded by a subcode of the $r$th-order Reed-Muller code ${\rm R}(r, m)$.
%Precisely,
$\dbu_0 = ((-1)^{b_0 ^{(0)}}, (-1)^{b_0 ^{(1)}},\cdots, (-1)^{b_0 ^{(2^m-1)}})$
where $\bbu_0 = (b_0 ^{(0)}, b_0 ^{(1)}, \cdots, b_0 ^{(2^m-1)})$ is a codeword
of a $(2^m, W)$ Reed-Muller subcode.
%In this section,
%We investigate the PAPR of the MC-CDMA signals where
%a variety of Reed-Muller subcodes are used for $\bbu_0$.
%First of all, we show that
%the first-order Reed-Muller provides a constant envelope power of PAPR $=1$
%for coded MC-CDMA signals in $K=L$ and $N=1$.
%We then introduce a recursive construction of a coding scheme
%which is a subcode of the $r$th order Reed-Muller code $R(r, m)$.
%We also show that the MC-CDMA signals employing the recursive coding schemes have the bounded PAPR.
%As specific examples, we introduce several codes from the recursive construction,
%and analyze the PAPRs of MC-CDMA signals employing the codes.
In this section, we denote $\bbu_0 = (b_0, b_1, \cdots, b_{2^m-1})$ for simplicity.

\subsection{The first-order Reed-Muller code}
Let $\bbu_0 = (b_0, b_1, \cdots, b_{2^m-1})$ be a codeword of the first-order Reed-Muller code ${\cal B}_1 ^{(m)} = {\rm R}(1, m)$.
When it is employed as a $(K, W)$ coding scheme in a coded MC-CDMA,
the dimension is $W = m+1$ and the codeword length is $K=2^m$.
Each codeword is associated with a Boolean function of % $b_1(x_0, \cdots, x_{m-1})$ given by %~\cite{Mac:ECC}
\begin{equation}\label{eq:bool_1}
b_1(x_0, \cdots, x_{m-1}) = \sum_{i=0} ^{m-1} v_i x_i +e, \quad v_i, e \in \{ 0, 1 \}
\end{equation}
where the addition is computed modulo-$2$.
%Then, each coded bit $b_k$ is given by
%\begin{equation}\label{eq:b_k}
%b_k =  b_1(k_0, \cdots, k_{m-1}) = \sum_{i=0} ^{m-1} v_i k_i +e
%\end{equation}
%where $k = \sum_{i=0} ^{m-1} k_i 2^i, \ k_i \in \{ 0, 1 \}$.
The PAPR of the MC-CDMA signal encoded by a codeword in ${\cal B}_1 ^{(m)}$ is determined in the following.

\vspace{0.1in}
\begin{thr}\label{th:papr_1}
With $K=L=2^m$ and $N=1$,
let $s_0(t)$ be a Walsh-Hadamard spread MC-CDMA signal
%in (\ref{eq:s_t_0}) where it is
encoded by $\bbu_0 = (b_0, b_1, \cdots, b_{2^m-1}) \in {\cal B}_1 ^{(m)} = R(1, m)$.
%where the dimension of ${\cal B}_1 ^{(m)}$ is $W=m+1$.
Then, the PAPR of $s_0(t)$ is % determined by
\[
{\rm PAPR} (s_0(t)) = 1.
\]
\end{thr}
\vspace{0.1in}

\noindent \emph{Proof.}
From Theorem~\ref{th:matrix_s}, the associated polynomial of $s_0(t)$ % in (\ref{eq:s_t_0})
is given by
$S_0(z) = \sqrt{\frac{w}{2^m}} \cdot \widehat{\dbu}_0 \cdot \zbu $ where
$\widehat{\dbu}_0 $ is
the Walsh-Hadamard transform of $\dbu_0$, % given by
and $\zbu = (1, z, z^2, \cdots, z^{2^m-1})^T$.
By definition,
$\widehat{\dbu}_0 = (\widehat{d}_0, \widehat{d}_1, \cdots, \widehat{d}_{2^m-1})$ where
\[
\widehat{d}_l = \frac{1}{\sqrt{2^m} }\sum_{k=0} ^{2^m-1} (-1)^{b_k + \sum_{i=0} ^{m-1} l_i k_i}
\quad \mbox{where } l = \sum_{i=0} ^{m-1} l_i 2^i, \ k = \sum_{i=0} ^{m-1} k_i 2^i.
\]
%where $$ and $$.
%is a row index of $\Hbu_{2^m}$.
In (\ref{eq:bool_1}), $ b_k = b_1(k_0, \cdots, k_{m-1}), \ 0 \leq k \leq 2^m-1$.
From (\ref{eq:exp_sum}),
\begin{equation*}
\begin{split}
\widehat{d}_l = & \frac{1}{\sqrt{2^m} } \sum_{k=0} ^{2^m-1} (-1)^{\sum_{i=0} ^{m-1} v_i k_i +e + \sum_{i=0} ^{m-1} l_i k_i}
= \frac{1}{\sqrt{2^m} } \sum_{k=0} ^{2^m-1} (-1)^{\sum_{i=0} ^{m-1} (v_i+l_i) k_i +e } \\
= & \left \{ \begin{array}{ll} \pm \sqrt{2^m}, & \quad \mbox{ if } l=v \\
0, & \quad \mbox{ otherwise} \end{array} \right.
\end{split}
\end{equation*}
where
$v = \sum_{i=0} ^{m-1} v_i 2^i$ for given $v_i$'s.
Therefore, %the associated polynomial of MC-CDMA signal $s_0(t)$ is therefore
$S_0(z) = \sqrt{\frac{w}{2^m}} \cdot \widehat{\dbu}_0 \cdot \zbu = \pm \sqrt{\frac{w}{2^m}} \cdot\sqrt{2^m} \cdot z^v
= \pm \sqrt{w} \cdot z^v$. % where $\zbu = (1, z, z^2, \cdots, z^{2^m-1})^T$.
For any $v$,
the PAPR of $s_0(t)$ is therefore % from (\ref{eq:S_z_papr}) % given by
\[
{\rm PAPR}(s_0(t)) = \frac{\max_{|z|=1} |S_0(z)|^2}{w} = \frac{w \cdot |z^{2v}|}{w} = 1.
\]
\qed
\vspace{0.1in}

From Theorem~\ref{th:papr_1}, we see that the first-order Reed-Muller code is a simple and effective coding scheme
that provides the uniform power for the coded MC-CDMA signals.
However, it has a relatively low code rate $R_1 = \frac{m+1}{2^m}$, which vanishes as the code length increases.
Therefore, we need to develop high-rate coding schemes at the expense of the PAPR increases.

\subsection{Recursive construction}
From a seed pair of codes,
we present how to recursively construct a new code using the associated Boolean functions.
We also analyze the PAPR of the coded MC-CDMA signals.
%which is a main result of this paper.
%Note that we study the PAPR properties for fully-loaded, Reed-Muller coded, and Walsh-Hadamard spread MC-CDMA signals,
%where $K=L=2^m$ and $N=1$.
%where the resultant code has the average of the code rates
%of the seed pair, while its length is twice and
%at the cost of the PAPR increase.

\vspace{0.1in}
\begin{thr}\label{th:recur}
Let $f$ and $g$ be Boolean functions of $(m-1)$ variables, where
$\bbu_f \in {\cal F}$ and $\bbu_g \in {\cal G}$ are the codewords of length $2^{m-1}$
associated with $f$ and $g$, respectively.
Assume that the code rate of each code % ${\cal F}$ and ${\cal G}$
is $R_f$ and $R_g$, respectively.
Let $s_f(t)$ and $s_g(t)$ be the coded MC-CDMA signals encoded by $\bbu_f$ and $\bbu_g$, respectively,
each of which has a form of (\ref{eq:s_t_0}) where $K=L=2^{m-1}$ and $N=1$.
Assume that each signal has the PAPR of at most $P$.

Consider a Boolean function $b$ of $m$ variables
defined by
\begin{equation}\label{eq:recur}
b(x_0, \cdots, x_{m-1}) = (1+x_{m-1}) \cdot f(x_0, \cdots, x_{m-2}) + x_{m-1} \cdot g(x_0, \cdots, x_{m-2}).
\end{equation}
Then, a codeword $\bbu_0$ of length $2^m$ associated with $b $ has the code rate $R = \frac{R_f + R_g}{2}$.
Let $s_0(t)$ be a coded MC-CDMA signal of (\ref{eq:s_t_0})
encoded by $\bbu_0$, where $K=L=2^m$ and $N=1$. %, associated with $b(x_0, \cdots, x_{m-1}) $.
Then, the PAPR of $s_0(t)$ is % given by
\begin{equation*}\label{eq:recur_papr}
{\rm PAPR} (s_0(t)) \leq 2P.
\end{equation*}
\end{thr}
\vspace{0.1in}

\noindent \emph{Proof.}
Obviously, $| {\cal F} | = 2^{2^{m-1} \cdot R_f}$ and
$| {\cal G} | = 2^{2^{m-1} \cdot R_g}$, respectively.
Thus, the number of codewords $\bbu_0$ is $ | {\cal F} |\cdot | {\cal G} | = 2^{2^{m-1} \cdot (R_f+R_g)}$ and the code rate is % obviously
$R = \frac{2^{m-1} \cdot (R_f+R_g)}{2^m} = \frac{R_f + R_g}{2}$.
In particular, if ${\cal F} = {\cal G}$, then we keep the code rate $R=R_f=R_g$ while the codeword length doubles.

%Let $\dbu_f = ((-1)^{b_{f, 0}}, \cdots, (-1)^{b_{f, 2^{m-1}-1}})$
%and $\dbu_g = ((-1)^{b_{g, 0}}, \cdots, (-1)^{b_{g, 2^{m-1}-1}})$,
%where $\bbu_f = (b_{f, 0}, \cdots, b_{f, 2^{m-1}-1})$ and $\bbu_g = (b_{g, 0}, \cdots, b_{g, 2^{m-1}-1})$.
Let $\dbu_f$ and $\dbu_g$ be the BPSK modulation outputs of length $2^{m-1}$ from $\bbu_f$ and $\bbu_g$, respectively.
From (\ref{eq:recur}),
it is straightforward that $\bbu_0 = (\bbu_f \ | \ \bbu_g)$ and $\dbu_0 = (\dbu_f \ | \ \dbu_g)$,
where `$|$' denotes a concatenation.
Then, the associated polynomial $S_0(z)$ is determined by
\begin{equation}\label{eq:recur_S_z}
\begin{split}
S_0(z) & = \sqrt{\frac{w}{2^m}} \cdot \dbu_0 \cdot \Hbu_{2^m} \cdot \zbu \\
& =  \sqrt{\frac{w}{2^m}} \cdot (\dbu_f \ | \ \dbu_g) \cdot \frac{1}{\sqrt{2}}
\begin{bmatrix}
\Hbu_{2^{m-1}} & \Hbu_{2^{m-1}} \\
\Hbu_{2^{m-1}} & - \Hbu_{2^{m-1}} \\
\end{bmatrix}
\cdot
\begin{bmatrix}
\zbu_1 \\
z^{2^{m-1}} \cdot \zbu_1 \\
\end{bmatrix} \\
& =  \sqrt{\frac{w}{2^m}} \cdot (\dbu_f \ | \ \dbu_g) \cdot
\begin{bmatrix}
\Hbu_{2^{m-1}} \cdot \zbu_1 \cdot \phi_{m-1} \\
\Hbu_{2^{m-1}} \cdot \zbu_1 \cdot \theta_{m-1} \\
\end{bmatrix} \\
\end{split}
\end{equation}
where $\zbu = (1, z, z^2, \cdots, z^{2^m-1})^T$, $\zbu_1 = (1, z, z^2, \cdots, z^{2^{m-1}-1})^T$,
$\phi_{m-1} = \frac{1+z^{2^{m-1}}}{\sqrt{2}}$, and $\theta_{m-1} = \frac{1-z^{2^{m-1}}}{\sqrt{2}}$.
Let $B_f(z) =  \sqrt{\frac{w}{2^m}} \cdot \dbu_f \cdot \Hbu_{2^{m-1}} \cdot \zbu_1$
and $B_g(z) =  \sqrt{\frac{w}{2^m}} \cdot \dbu_g \cdot \Hbu_{2^{m-1}} \cdot \zbu_1$.
Then, (\ref{eq:recur_S_z}) becomes
\[
S_0(z) = B_f(z)\cdot \phi_{m-1} + B_g(z) \cdot \theta_{m-1}.
\]
Replacing the Boolean function $g$ by $g+1$ leads us to the change of the above associated polynomial to $S_0 '(z)$, i.e.,
\[
S_0 '(z) = B_f(z)\cdot \phi_{m-1} - B_g(z) \cdot \theta_{m-1}.
\]
Then,
\begin{equation}\label{eq:recur_S_z_sum}
|S_0(z)|^2 + |S_0 '(z)|^2 = 2 \cdot \left(|B_f(z)|^2 \cdot |\phi_{m-1}|^2 + |B_g(z)|^2 \cdot |\theta_{m-1}|^2 \right).
\end{equation}
If $s_f(t)$ and $s_g(t)$ have the PAPR of at most $P$, then Corollary~\ref{co:S_z_sqr}
%(\ref{eq:dCz})
implies
$\left| \dbu_f \cdot \Hbu_{2^{m-1}} \cdot \zbu_1 \right|^2 \leq 2^{m-1} P$
and $\left| \dbu_g \cdot \Hbu_{2^{m-1}} \cdot \zbu_1 \right|^2 \leq 2^{m-1} P$, respectively.
Thus,
\begin{equation}\label{eq:B_f_B_g}
|B_f(z)|^2 \leq \frac{w}{2^m} \cdot 2^{m-1} \cdot P = \frac{w}{2} \cdot P, \quad |B_g(z)|^2 \leq \frac{w}{2}  \cdot P.
\end{equation}
Therefore, from (\ref{eq:recur_S_z_sum}) and (\ref{eq:B_f_B_g}),
\begin{equation*}
\begin{split}
|S_0(z)|^2 + |S_0 '(z)|^2 & \leq 2 \cdot \max \left(|B_f(z)|^2 ,  |B_g(z)|^2 \right) \cdot (|\phi_{m-1}|^2 + |\theta_{m-1}|^2) \\
& = 2 \cdot\frac{w}{2}  \cdot P \cdot 2 = w \cdot 2P
\end{split}
\end{equation*}
where $|\phi_{m-1}|^2 + |\theta_{m-1}|^2 = 2$ from the definition of $\phi_{m-1}$ and $\theta_{m-1}$.
Thus, the PAPR of $s_0(t)$ is % given by
\[
{\rm PAPR} (s_0(t)) = \frac{\max_{|z|=1} |S_0(z)|^2}{w} \leq 2P.
\]
\qed
\vspace{0.1in}

%From the recursive construction described in Theorem~\ref{th:recur},
%we are able to construct a code of length $2^m$ by \emph{concatenating} codewords from the seed code pair.
%%${\cal B}_f ^{(m-1)}$ and ${\cal B}_g ^{(m-1)}$.
%In fact, the concatenation allows any construction of a code of length $2^m$, code rate $R$, and the PAPR of at most $2P$,
%from a seed code of length $2^{m-1}$, code rate $R$, and the PAPR of at most $P$.
%A similar technique has been discussed in~\cite{Pat:Codes} for multicode CDMA.

%In addition to the coding schemes in Constructions~\ref{cst:B2} and~\ref{cst:B3},
%we may establish a variety of coding schemes from the recursive construction in Theorem~\ref{th:recur}
%by employing a known code as a seed pair of $f$ and $g$.
The recursive construction of a Boolean function has been originally discussed in~\cite{Pat:Codes} for
the PAPR of multicode CDMA.
In Theorem~\ref{th:recur}, we showed that the construction of (\ref{eq:recur})
also provides the bounded PAPR for multicarrier CDMA.
In general, if there is a seed code ${\cal B}_{r-1} ^{(m-1)}$ of length $2^{m-1}$ and size $|{\cal B}_{r-1 } ^{(m-1)} |$,
then we can construct a new code ${\cal B}_{r} ^{(m)}$ of length $2^m$ and size
$|{\cal B}_{r} ^{(m)}|=|{\cal B}_{r-1} ^{(m-1)}|^2$
%from the recursive construction described in Theorem~\ref{th:recur}
by \emph{concatenating} a pair of codewords from the seed. % as described in Theorem~\ref{th:recur}.
If the PAPR of each coded MC-CDMA signal for the seed ${\cal B}_{r-1} ^{(m-1)}$ is at most $P$,
then each coded MC-CDMA signal encoded by the new code ${\cal B}_{r} ^{(m)}$ provides the PAPR of at most $2P$.
%from Theorem~\ref{th:recur}.
If each codeword in ${\cal B}_{r-1} ^{(m-1)}$ is associated with a Boolean function of degree at most $r-1$,
then ${\cal B}_{r} ^{(m)}$ is a subcode of ${\rm R}(r, m)$
defined by a Boolean function of degree $r$,
where the minimum Hamming distance of ${\cal B}_{r} ^{(m)}$ is at least $2^{m-r}$.

Construction~\ref{cst:gen_code} summarizes a recursive code construction % in a recursive way
for the application to MC-CDMA.

\vspace{0.1in}
\begin{const}\label{cst:gen_code}
For positive integers $r $ and $m$, $2 \leq r \leq m$,
let $b_1(x_0, \cdots, x_{m-r}) = \sum_{i=0} ^{m-r} v_i x_i + e$
and $b_1 ' (x_0, \cdots, x_{m-r}) = \sum_{i=0} ^{m-r} v_i' x_i + e'$,
where $v_i, v_i',e, e' \in \{ 0, 1 \}$.
Starting with $b_1$ and
$b_1 '$,
the Boolean function $b_r (x_0, \cdots, x_{m-1})$ of degree $r$ is constructed
by the $(r-1)$ successive recursions of
%is recursively constructed by %(\ref{eq:recur}). % where $b^{(r-1)}(x_0, \cdots, x_{m-2})$ is employed as $f$ and $g$.
\begin{equation}\label{eq:b_r_bool}
\begin{split}
b_{\tilde{ r} } (x_0, \cdots, x_{m-r+\tilde{r}-1})  = & (1+x_{m-r+\tilde{r}-1}) \cdot b_{\tilde{r}-1} (x_0, \cdots, x_{m-r+\tilde{r}-2}) \\
& + x_{m-r+\tilde{r}-1} \cdot b_{\tilde{r}-1} ' (x_0, \cdots, x_{m-r+\tilde{r}-2}) %  2 \leq \tilde{ r} \leq r
\end{split}
\end{equation}
while $\tilde{r}$ runs through $2$ to $r$.
In (\ref{eq:b_r_bool}), the Boolean function $b_{\tilde{r}-1} '$
has the same form as $b_{\tilde{r}-1}$, but may have different coefficients. %, as in $b_1'$.
% are associated with codewords in ${\cal B}_{r-1} ^{(m-1)}$.
%we can generate a Boolean function $b_r (x_0, \cdots, x_{m-1})$ by the $(r-1)$ successive recursions.
%that is associated with the Boolean function $b^{(r)}$.
Let $\bbu_0 = (b_0, b_1, \cdots, b_{2^m-1})$ be a codeword of ${\cal B}_r ^{(m)} \subset {\rm R}(r, m)$,
associated with a Boolean function of $b_r$. % of degree $r$.
Then,
${\cal B}_r ^{(m)}$ has total $2^{(m-r+2) \cdot 2^{r-1}}$ codewords through the $(r-1)$ recursions.
In a Walsh-Hadamard spread MC-CDMA with $K=L=2^m$ and $N=1$,
the PAPR of $s_0(t)$ encoded by $\bbu_0$ % a codeword in ${\cal B}_r ^{(m)}$
is at most $2^{r-1}$ from Theorems~\ref{th:papr_1} and~\ref{th:recur}.

Finally, %for $2 \leq r \leq m$,
the code parameters of ${\cal B}_r ^{(m)}$ %and % the dimension, code length, code rate minimum distance $d_{\min}$, and
%the PAPR of the coded MC-CDMA signal $s_0(t)$ corresponding to a codeword in ${\cal B}_r ^{(m)}$ for $2 \leq r \leq m$
are summarized as follows.
\begin{itemize}
\item[$-$] Dimension $W = 2^{r-1}\cdot (m-r+2)$ and code length $K = 2^m$,
\item[$-$] Code rate $R_r = \frac{2^{r-1}\cdot (m-r+2)}{2^m} = 2^{r-1-m} \cdot (m-r+2)$,
\item[$-$] Minimum Hamming distance $ \geq 2^{m-r}$,
\item[$-$] ${\rm PAPR}(s_0(t)) \leq 2^{r-1}$.
\end{itemize}
%Then, the dimension of ${\cal B}_r ^{(m)}$ is $W = 2^{r-1}\cdot (m-r+2)$,
%and thus the code rate $R_r = \frac{2^{r-1}\cdot (m-r+2)}{2^m} = 2^{r-1-m} \cdot (m-r+2)$.
%Since ${\cal B}_r ^{(m)} $ is a subcode of ${\rm R}(r, m)$,
%the minimum distance is at least $2^{m-r}$.
%Finally, the PAPR of the coded MC-CDMA signal $s_0(t)$
%corresponding to a codeword in ${\cal B}_r ^{(m)}$
%is bounded by
%${\rm PAPR}(s_0(t)) \leq 2^{r-1}$.
\end{const}
\vspace{0.1in}

First of all, we present a specific code example ${\cal B}_2 ^{(m)}$ of length $2^m$
through a single recursion,
where a pair of codewords in ${\rm R}(1, m-1)$ is employed as the seed.
%Both schemes provide higher code rates than the first-order Reed-Muller codes at the cost of PAPR increase.

\vspace{0.1in}
\begin{const}\label{cst:B2}
Let $\bbu_0 = (b_0, b_1, \cdots, b_{2^m-1})$ be a codeword of ${\cal B}_2 ^{(m)} \subset {\rm R}(2, m)$
that is associated with a Boolean function $b_2$ defined by
\begin{equation}\label{eq:bool_2}
b_2(x_0, \cdots, x_{m-1}) = \sum_{i=0} ^{m-1} v_i x_i + x_{m-1} \cdot \sum_{i=0} ^{m-2} v_i ' x_i + e,
\quad v_i, v_i ', e \in \{0, 1\}.
\end{equation}
In a Walsh-Hadamard spread MC-CDMA with $K=L=2^m$ and $N=1$,
the code parameters including
the PAPR of a coded MC-CDMA signal $s_0(t)$ encoded by $\bbu_0$ % a codeword in ${\cal B}_2 ^{(m)}$
are summarized as follows.
\begin{itemize}
\item[$-$] Dimension $W = 2m$ and code length $K = 2^m$,
\item[$-$] Code rate $R_2 = \frac{m}{2^{m-1}}$,
\item[$-$] Minimum Hamming distance $ \geq 2^{m-2}$,
\item[$-$] ${\rm PAPR}(s_0(t)) \leq 2$.
\end{itemize}
\end{const}
\vspace{0.1in}

%\begin{thr}\label{th:B2_papr}
%Let $s_0(t)$ be the coded MC-CDMA signal corresponding to a codeword $\bbu \in {\cal B}_2 ^{(m)}$
%in Construction~\ref{cst:B2}.
%Then, the PAPR of $s_0(t)$ is bounded by
%\begin{equation}\label{eq:papr_2}
%{\rm PAPR}(s_0(t)) \leq 2.
%\end{equation}
%\end{thr}

%\noindent \emph{Proof.}
%Let $f$ and $g$ be the Boolean functions of (\ref{eq:bool_1})
%generating the first-order Reed-Muller code of length $2^{m-1}$.
With a single recursion ($r = \tilde{r} =2$) in Construction~\ref{cst:gen_code},
%If we employ the Boolean functions in (\ref{eq:bool_1}) for a seed pair in the recursion of (\ref{eq:recur}),
(\ref{eq:bool_2}) is straightforward by
\begin{equation*}
\begin{split}
b_2(x_0, \cdots, x_{m-1}) & = (1+x_{m-1}) \cdot \left(\sum_{i=0} ^{m-2} v_i x_i +e \right)
+ x_{m-1} \cdot  \left(\sum_{i=0} ^{m-2} v_i '' x_i +e ' \right) \\
& = \sum_{i=0} ^{m-2} v_i x_i + (e+e') \cdot x_{m-1} + x_{m-1} \cdot  \sum_{i=0} ^{m-2} (v_i + v_i '') \cdot x_i  + e \\
& =  \sum_{i=0} ^{m-1} v_i x_i + x_{m-1} \cdot  \sum_{i=0} ^{m-2}  v_i ' x_i +e
\end{split}
\end{equation*}
where $v_{m-1} = e+e'$ and $v_i ' =v_i + v_i ''$.
%Since $f$ and $g$ generate the codewords in ${\cal B}_1$,
%(\ref{eq:papr_2}) is immediate from
%Theorems~\ref{th:papr_1} and \ref{th:recur},
%\qed

\vspace{0.1in}
\begin{rem}\label{rem:B2}
By generalizing (\ref{eq:bool_2}) to
\begin{equation*}\label{eq:b2_r}
g_2(x_0, \cdots, x_{m-1}) = \sum_{i=0} ^{m-1} v_i x_i + x_{\gamma} \cdot \sum_{i=0} ^{\gamma} v_i ' x_i + e, \quad
1 \leq \gamma \leq m-1,
\end{equation*}
we obtain a code ${\cal GB}_2 ^{(m)}$ associated with $g_2$.
Obviously, the coding scheme ${\cal B}_2 ^{(m)}$ in Construction~\ref{cst:B2} is a special case of
${\cal GB}_2 ^{(m)}$ for $\gamma=m-1$.
It is not so hard to prove that a coded MC-CDMA signal $s_0(t)$ encoded by a codeword % $\bbu_\gamma$,
associated with $g_2$ has the PAPR of at most $2$ for any $\gamma$.
While $\gamma$ runs through $1$ to $m-1$, we have
$ 2^{m+1} \cdot (2^{m-1} + 2^{m-2}-1 + 2^{m-3}-1 + \cdots + 2-1)
= 2^{m+1} \cdot (2^m-m) $
distinct codewords in ${\cal GB}_2 ^{(m)}$, % from $g_2$,
more than the number of codewords in ${\cal B}_2 ^{(m)}$.
If we compare the code rates of ${\cal GB}_2 ^{(m)}$ and ${\cal B}_2 ^{(m)}$, however,
the code rate difference of
\[
\frac{\log_2 ( 2^{m+1} \cdot (2^m-m) ) -2m}{2^m}
= \frac{\log_2 (2^m-m ) +1-m}{2^m} < \frac{1}{2^m}
\]
is very small and approaches to $0$ as $m$ increases.
Meanwhile, the encoding and the decoding complexities of the generalized coding scheme ${\cal GB}_2 ^{(m)}$ are obviously larger
than those of ${\cal B}_2 ^{(m)}$.
For the little contribution to the code rate and the increase of
the complexities from ${\cal GB}_2 ^{(m)}$, we therefore consider ${\cal B}_2 ^{(m)}$ as our coding scheme for low PAPR.
\end{rem}
\vspace{0.1in}

By employing the coding scheme ${\cal B}_2 ^{(m)}$, the Walsh-Hadamard spread MC-CDMA system with $K=L=2^m$ and $N=1$
is able to support maximum $2m$ users % (downlink)
or $2m$ information bits from a single user in an OFDM symbol, %(uplink) %and Walsh-Hadamard spreading sequences of length $2^m$,
providing the PAPR of at most $2$.

%Next, we employ the Boolean function $b_2$ in (\ref{eq:bool_2}) as a seed in the recursive construction
%to present another coding scheme ${\cal B}_3 ^{(m)}$ with the PAPR of at most $4$.

\vspace{0.1in}
\begin{const}\label{cst:B3}
Let a Boolean function $b_3$ be defined by
\begin{equation}\label{eq:bool_3}
b_3(x_0, \cdots, x_{m-1}) = \sum_{i=0} ^{m-1} v_i x_i + x_{m-1}x_{m-2} \cdot \sum_{i=0} ^{m-3} v_i ' x_i
+  x_{m-1} \cdot \sum_{i=0} ^{m-2} v_i '' x_i +  x_{m-2} \cdot \sum_{i=0} ^{m-3} v_i ''' x_i  + e
\end{equation}
where $v_i, v_i ', v_i'', v_i ''', e \in \{0, 1\}$.
Let $\bbu_0 = (b_0, b_1, \cdots, b_{2^m-1})$ be a codeword of ${\cal B}_3 ^{(m)} \subset {\rm R}(3, m)$
that is associated with $b_3$.
%From $v_i, v_i ', v_i'', v_i '''$ and $ e$,
In a Walsh-Hadamard spread MC-CDMA with $K=L=2^m$ and $N=1$,
the code parameters including
the PAPR of a coded MC-CDMA signal $s_0(t)$ encoded by $\bbu_0$
are summarized as
\begin{itemize}
\item[$-$] Dimension $W = 4(m-1)$ and code length $K = 2^m$,
\item[$-$] Code rate $R_3 =\frac{m-1}{2^{m-2}}$,
\item[$-$] Minimum Hamming distance $ \geq 2^{m-3}$,
\item[$-$] ${\rm PAPR}(s_0(t)) \leq 4$.
%The dimension of ${\cal B}_3 ^{(m)}$ is $W = 4m-4$, and thus the code rate $R = \frac{m-1}{2^{m-2}}$.
%Since ${\cal B}_3 ^{(m)}$ is a subcode of ${\rm R}(3, m)$,
%the minimum distance is at least $2^{m-3}$.
\end{itemize}
\end{const}
\vspace{0.1in}

%\begin{thr}\label{th:B3_papr}
%Let $s_0(t)$ be the coded MC-CDMA signal corresponding to a codeword $\bbu \in {\cal B}_3 ^{(m)}$
%in Construction~\ref{cst:B3}.
%Then, the PAPR of $s_0(t)$ is bounded by
%\begin{equation}\label{eq:papr_3}
%{\rm PAPR}(s_0(t)) \leq 4.
%\end{equation}
%\end{thr}

%\noindent \emph{Proof.}
Similar to Construction~\ref{cst:B2},
%the proof of Theorem~\ref{th:B2_papr},
the Boolean function $b_3$ is immediate from
the twice recursions of (\ref{eq:recur}) for $\tilde{r} = 2$ and $3$ with $r=3$.
%(\ref{eq:recur}), where $f$ and $g$ are the Boolean functions of (\ref{eq:bool_2}).
%and thus the PAPR of
%(\ref{eq:papr_2}) is immediate
%from Theorems~\ref{th:recur} and \ref{th:B2_papr}.
%\qed
%\vspace{0.1in}
With the coding scheme ${\cal B}_3 ^{(m)}$, the Walsh-Hadamard spread MC-CDMA system with $K=L=2^m$ and $N=1$
is able to support maximum $(4m-4)$ users %(downlink)
or $(4m-4)$ information bits from a single user in an OFDM symbol, %(uplink) % and Walsh-Hadamard spreading sequences of length $2^m$,
providing the PAPR of at most $4$.

%\vspace{0.1in}
%\begin{rem}\label{rem:B3}
%If we apply the Boolean function (\ref{eq:b2_r}) in Remark~\ref{rem:B2} % where $1 \leq \gamma \leq m-2$,
%for a code pair $f$ and $g$ in the recursive construction, then %in (\ref{eq:recur}), then
%the resultant code ${\cal GB}_3 ^{(m)}$ has more distinct codewords than ${\cal B}_3 ^{(m)}$, while it
%provides the PAPR $\leq 4$ for the coded MC-CDMA signals.
%While $\gamma$ runs through $1$ to $m-2$ in $f$ and $g$, the number of distinct codewords in ${\cal GB}_3 ^{(m)}$ is
%$\left(2^m \cdot (2^{m-1} - m+1) \right)^2 = 2^{2m} \cdot (2^{m-1}-m+1)^2$.
%Compared to ${\cal B}_3 ^{(m)}$, the code rate increase is
%\[
%\frac{\log_2 \left( 2^{2m} \cdot (2^{m-1}-m+1)^2 \right) -4m+4}{2^m}
%= \frac{2 \log_2 (2^{m-1}-m+1)  -2m+4}{2^m} < \frac{1}{2^{m-1}}.
%\]
%For the similar concerns regarding the code rates and the complexities
%mentioned in Remark~\ref{rem:B2},
%we consider ${\cal B}_3 ^{(m)}$ as our coding scheme for low PAPR.
%%due to
%%the little contribution to the code rate and the increase of
%%encoding and decoding complexity.
%\end{rem}
%\vspace{0.1in}

%In conclusion, we may construct a high rate code from the recursive construction at the cost of
%the PAPR increase.

Table~\ref{tb:gen_recur} lists the parameters of ${\cal B}_r ^{(m)} $ of length $2^m$ from Construction~\ref{cst:gen_code},
where $1 \leq r \leq 5$ and $m \geq r$.
It elucidates a connection between the code rates of ${\cal B}_r ^{(m)} $ and the maximum PAPR of MC-CDMA signals
encoded by ${\cal B}_r ^{(m)} $.
%Note that we construct ${\cal B}_r ^{(m)} $, $r \geq 2$, by applying the recursive construction to ${\cal B}_{r-1}$ of length $2^{m-1}$.
%As $r$ increases, we experience more recursions to obtain a high rate code.
%Each code has different dimension, code rate, degree of Boolean function, minimum distance, and maximum PAPR
%for the coded MC-CDMA signals.
Obviously, we obtain a high rate Reed-Muller subcode at the cost of high PAPR for the coded MC-CDMA signals.

\begin{table}
\fontsize{8}{10pt}\selectfont
\caption{The parameters of ${\cal B}_r ^{(m)}$ of length $2^m$ for some $r$'s}
\centering
\begin{tabular}{|c|c|c|c|c|}
\hline
Code &  Dimension $W$   &   Degree $r$        &   Minimum    &  Maximum PAPR     \\
             &     &   of a Boolean function           &    Hamming distance    &  (for $K=L=2^m$ and $N=1$) \\
\hline
${\cal B}_1$  &  $m+1$        & $1$                    &   $\geq 2^{m-1}$     &   $1$            \\
${\cal B}_2$  & $2m$        &    $2$                    &   $\geq 2^{m-2}$     &   $2$            \\
${\cal B}_3$  & $4(m-1)$       &   $3$                    &   $\geq 2^{m-3}$     &   $4$         \\
${\cal B}_4$ & $8(m-2)$       &   $4$                    &   $\geq 2^{m-4}$     &   $8$              \\
${\cal B}_5$ & $16(m-3)$       &   $5$                    &   $\geq 2^{m-5}$     &   $16$           \\
\hline
\end{tabular}
\label{tb:gen_recur}
\end{table}

\vspace{0.1in}
\begin{rem}\label{rem:RM_map}
%Table~\ref{tb:gen_recur} shows a relation between code rates and maximum PAPRs
%of the coded MC-CDMA.
%In general, the recursive construction presents a code ${\cal B}_r$ of length $2^m$
%associated with a Boolean function of degree $r$, $1 \leq r \leq m$,
%where the code rate and the PAPR of $s(t)$ are given by
%$ R_r = \frac{2^{r-1}\cdot (m-r+2)}{2^m} = 2^{r-1-m} \cdot (m-r+2)$ and
%${\rm PAPR}(s(t)) \leq 2^{r-1}$, respectively.
In Construction~\ref{cst:gen_code}, if $r=m$, then
$R_r = 1$ and ${\rm PAPR}(s_0(t)) \leq 2^{m-1}$ in ${\cal B}_r ^{(m)}$.
In other words, if we apply a trivial $m$th-order Reed-Muller code of length $2^m$ and code rate $1.0$,
or equivalently apply a \emph{Reed-Muller mapping} to the information bits of length $2^m$,
then the PAPR of the corresponding MC-CDMA signals is bounded by $2^{m-1}$.
The Reed-Muller mapping may be useful for the applications
of small spreading factors, e.g., $m=2,3,4$, requiring low PAPR and no rate loss.
However, if the spreading factor is large, we may have the problems of the high PAPR and the large complexity of demapping at the receiver.
\end{rem}

\subsection{Golay complementary sequences}
Golay complementary sequences~\cite{Golay} provide the bounded PAPR $\leq 2$ for transmitted OFDM signals when they are employed
as a coding scheme in an OFDM system.
From~\cite{DavJed:GDJ}, it is well known that each binary Golay complementary sequence of length $2^m$
is equivalent to the second-order
coset of the first-order Reed-Muller code, where the associated Boolean function is defined by
\begin{equation}\label{eq:GDJ}
b_c(x_0, \cdots, x_{m-1}) = \sum_{i=0} ^{m-2} x_{\pi(i)} x_{\pi(i+1)} + \sum_{i=0} ^{m-1} v_i x_i + e
\end{equation}
where $\pi$ is a permutation in $\{ 0, 1, \cdots, m-1 \}$
and $v_i, e \in \{0, 1\}$.
We now apply a binary code ${\cal B}_c ^{(m)}$ for coded MC-CDMA signals,
where each codeword is a Golay complementary sequence defined by $b_c$.

\vspace{0.1in}
\begin{const}\label{cst:B_c}
Let $\bbu_0 = (b_0, b_1, \cdots, b_{2^m-1})$ be a codeword of ${\cal B}_c ^{(m)} \subset {\rm R}(2, m)$
that is associated with a Boolean function $b_c$ in (\ref{eq:GDJ}).
Then, its code parameters
are summarized as~\cite{DavJed:GDJ}
\begin{itemize}
\item[$-$] Dimension $W =  m+\log_2 (m!)$ and code length $ = 2^m$,
\item[$-$] Code rate $R_c = \frac{m+ \log_2(m!)}{2^m}$,
\item[$-$] Minimum Hamming distance $ \geq 2^{m-2}$.
%\item[$-$] ${\rm PAPR}(s_0(t)) \leq 4$.
\end{itemize}
\end{const}
\vspace{0.1in}

Through a linear unitary transform (LUT), Parker and Tellambura~\cite{ParkTell:GDJ}
have implicitly investigated the PAPR of MC-CDMA signals
encoded by the Golay complementary sequences.
Therefore, Theorem~\ref{th:Bc_papr} is immediate from the application of their work in a Walsh-Hadamard spread MC-CDMA.

\vspace{0.1in}
\begin{thr}\cite{ParkTell:GDJ}\label{th:Bc_papr}
In a Walsh-Hadamard spread MC-CDMA with $K=L=2^m$ and $N=1$,
let $s_0(t)$ be the coded MC-CDMA signal
encoded by a codeword $\bbu_0 \in {\cal B}_c ^{(m)}$
in Construction~\ref{cst:B_c}.
Then, % the PAPR of $s_0(t)$ is bounded by
\begin{equation}\label{eq:papr_c}
{\rm PAPR}(s_0(t)) \leq 2^{m- \lfloor \frac{m}{2} \rfloor}.
\end{equation}
\end{thr}

\noindent \emph{Proof.}
From Lemma~\ref{lemma:Cz}, $\Tbu = \Hbu_{2^m} \cdot \zbu$ is a special case of a linear unitary transform (LUT)
described in Theorem 6 of~\cite{ParkTell:GDJ}, where
$\phi_t = \frac{1+z^{2^t}}{\sqrt{2}}$
and $\theta_t =  \frac{1-z^{2^t}}{\sqrt{2}}$.
%where
%$|\phi_t|^2 + |\theta_t|^2 = 2$.
Theorem~\ref{th:matrix_s} implies that
$s_0(t)$ is equivalent to a unitary transform of a modulated Golay complementary sequence $\dbu_0$ through $\Tbu$.
Therefore, the bounded PAPR of~(\ref{eq:papr_c}) is obvious from Corollary 6 in~\cite{ParkTell:GDJ}.
\qed
\vspace{0.1in}

\begin{rem}
Numerical experiments revealed that
the actual maximum PAPR of MC-CDMA signals encoded by Golay complementary sequences
is smaller than the upper bound predicted by Theorem~\ref{th:Bc_papr}.
This may be the case because the PAPR bound of Theorem~\ref{th:Bc_papr}
has been established by general $\phi_t$ and $\theta_t$
with the requirement of $|\phi_t|^2 + |\theta_t|^2 = 2$~\cite{ParkTell:GDJ}.
However, the Walsh-Hadamard spread MC-CDMA has the special values of
$\phi_t = \frac{1+z^{2^t}}{\sqrt{2}}$ and $\theta_t = \frac{1-z^{2^t}}{\sqrt{2}}$, respectively,
which may require the tighter bound on the PAPR than (\ref{eq:papr_c}).
Table~\ref{tb:Golay_papr} shows the PAPR comparison between the theoretical bound and the numerical results.
From the numerical results, we conjecture ${\rm PAPR}(s_0(t)) \leq 2^{\lfloor \frac{m+1}{3} \rfloor}$,
where the proof is left open.
\end{rem}
\vspace{0.1in}

\begin{table}
\fontsize{8}{10pt}\selectfont
\caption{The maximum PAPR of Walsh-Hadamard spread MC-CDMA signals encoded by Golay complementary sequences of length $2^m$
($K=L=2^m$, $N=1$)}
\centering
\begin{tabular}{|c|c|c|c|}
\hline
$m$    &   Code length &  Theoretical maximum (Theorem~\ref{th:Bc_papr})    & Actual maximum (Numerical experiments) \\
\hline
$3$    &    $8$        &   $ 4$                       &   $1.9654 \ (\leq 2) $              \\
$4$    &    $16$        &   $ 4$                       &   $1.9998 \ (\leq 2)$              \\
$5$    &    $32$        &   $ 8$                       &   $3.2184 \ (\leq 4)$              \\
$6$    &    $64$        &   $ 8$                       &   $3.8826 \ (\leq 4)$              \\
$7$    &    $128$        &   $16$                       &   $3.9930 \ (\leq 4)$              \\
$8$    &    $256$        &   $16$                       &   $5.8964 \ (\leq 8)$              \\
\hline
\end{tabular}
\label{tb:Golay_papr}
\end{table}

Table~\ref{tb:code_rate} compares the code rates of ${\cal B}_2 ^{(m)}$, ${\cal B}_3 ^{(m)}$, and ${\cal B}_c ^{(m)}$
for several $m$'s.
We see from the table that ${\cal B}_3 ^{(m)}$ is a proper coding scheme for a Walsh-Hadamard spread MC-CDMA
with a small spreading factor,
providing the bounded PAPR $\leq 4$ and the acceptable code rates.

\begin{table}
\fontsize{8}{10pt}\selectfont
\caption{The code rates of various codes of length $2^m$}
\centering
\begin{tabular}{|c|c|c|c|c|}
\hline
$m$    &   Code length &  ${\cal B}_2$                 & ${\cal B}_3$                &   ${\cal B}_c$  \\
       &               &  (Construction~\ref{cst:B2})  & (Construction~\ref{cst:B3}) &  (Construction~\ref{cst:B_c})  \\
\hline
$3$    &    $8$        &   $0.75$                      &   $1.0$                     &   $0.6981$  \\
$4$    &    $16$        &   $0.5$                      &   $0.75$                    &   $0.5366$  \\
$5$    &    $32$        &   $0.3125$                   &   $0.5$                     &   $0.3721$  \\
$6$    &    $64$        &   $0.1875$                   &   $0.3125$                  &   $0.2421$  \\
$7$    &    $128$        &   $0.1094$                  &   $0.1875$                  &   $0.1508$  \\
%$8$    &    $256$        &   $16$                       &   $8$              \\
\hline
\end{tabular}
\label{tb:code_rate}
\end{table}

\subsection{Encoding and decoding}
In Figure~\ref{fig:coded_sys},
the Reed-Muller subcode introduced in Construction~\ref{cst:gen_code}
is applied at each spreading process for encoding the information from a single or multiple users
in the Reed-Muller coded and Walsh-Hadamard spread MC-CDMA transmitter ($K=L=2^m$).
Precisely,
%a $w$-bit information block multiplexed from each user becomes the input to a $(2^m, W)$ Reed-Muller subcode, and
%if $w\leq W$, we then append zeros to the end of
%the original information block to produce
%a $W$-bit input data
%$\abu_0 = (a_0 ^{(0)}, \cdots, a_0 ^{(w-1)}, 0, \cdots, 0)$.
%Then,
a $(2^m, W)$ Reed-Muller subcode ${\cal B}_r ^{(m)}$
encodes a $W$-bit input data $\abu_n= (a_n ^{(0)}, \cdots, a_n ^{(w-1)}, 0, \cdots, 0)$ at the $n$th spreading process, $0 \leq n \leq N-1$,
to produce a codeword $\bbu_n$ of length $2^m$,
which goes through Walsh-Hadamard spreading, interleaving,
and IFFT in the sequel.

In the encoding process, % for ${\cal B}_r ^{(m)}, \ r \geq 2$,
the codeword $\bbu_n $ is obtained by
\[
\bbu_n = \abu_n \cdot \Gbu_r ^{(m)}
\]
where $\Gbu_r ^{(m)}$ is the $W \times 2^m$ generator matrix of % the $(2^m, W)$ Reed-Muller subcode
${\cal B}_r ^{(m)}$,
where $W=2^{r-1} (m-r+2)$. % from Construction~\ref{cst:gen_code}.
The recursion of Boolean functions in (\ref{eq:b_r_bool})
equivalently derives the recursion of generator matrices of
\begin{equation}\label{eq:G_r_org}
\Gbu_{\tilde{r}} ^{(m-r+\tilde{r})} =
\begin{bmatrix}
\Gbu_{\tilde{r}-1} ^{(m-r+\tilde{r}-1)}  & {\bf 0}  \\
{\bf 0} & \Gbu_{\tilde{r}-1} ^{(m-r+\tilde{r}-1)} \\
\end{bmatrix}, \quad 2 \leq \tilde{r} \leq r
\end{equation}
where ${\bf 0} = (0, \cdots, 0)$ of length $2^{m-r+\tilde{r}-1}$
and $\Gbu_{\tilde{r}} ^{(m-r+\tilde{r})}$ is a $2^{\tilde{r}-1}(m-r+2) \times 2^{m-r+\tilde{r}}$ matrix.
By elementary row operations, it is
equivalent to
\begin{equation}\label{eq:G_r}
\Gbu_{\tilde{r}} ^{(m-r+\tilde{r})} =
\begin{bmatrix}
\Gbu_{\tilde{r}-1} ^{(m-r+\tilde{r}-1)}  & \Gbu_{\tilde{r}-1} ^{(m-r+\tilde{r}-1)}  \\
{\bf 0} & \Gbu_{\tilde{r}-1} ^{(m-r+\tilde{r}-1)} \\
\end{bmatrix}, \quad 2 \leq \tilde{r} \leq r .
\end{equation}
%it is straightforward to show that
While $\tilde{r}$ runs through $2$ to $r$,
the generator matrix $\Gbu_r ^{(m)}$ is constructed by the $(r-1)$ recursions of (\ref{eq:G_r_org}) or (\ref{eq:G_r}),
%which generates the same code ${\cal B}_r ^{(m)}$.
where the initial matrix % generator matrix
$\Gbu_1 ^{(m-r+1)}$ is
the $(m-r+2) \times 2^{m-r+1}$ generator matrix of ${\rm R}(1, m-r+1)$ given by %~\cite{Mac:ECC}
\begin{equation*}\label{eq:G1}
\Gbu_1 ^{(m-r+1)} =
\begin{bmatrix}
1111 & 1111 &  \cdots  & 1111 & 1111 \\
0101 & 0101 &  \cdots  & 0101 & 0101 \\
0011 & 0011 &  \cdots  & 0011 & 0011 \\
0000 & 1111 &  \cdots  & 0000 & 1111 \\
\vdots &  \vdots &  & \vdots  & \vdots\\
0000 & 0000 &  \cdots  & 1111 & 1111 \\
\end{bmatrix} =
\begin{bmatrix}
{\bf 1} \\
{\xbu}_0 \\
{\xbu}_1 \\
{\xbu}_2 \\
\vdots \\
{\xbu}_{m-r}
\end{bmatrix}.
\end{equation*}
For the notations $\xbu_i$'s of the generator matrix of the first-order Reed-Muller codes, see~\cite{Mac:ECC}.

In particular, we are able to determine $\Gbu_2 ^{(m)}$ and $\Gbu_3 ^{(m)}$
directly from the Boolean expressions in (\ref{eq:bool_2}) and (\ref{eq:bool_3}), respectively.
The generator matrices of ${\cal B}_2$ and ${\cal B}_3$
are $2m \times 2^m$ and $(4m-4) \times 2^m$ matrices, respectively.
%each of which determines a subcode of $R(2, m)$ and $R(3, m)$, respectively.
%Recall the notations of $\xbu_i$ in (\ref{eq:G1}).
With each $\xbu_i$ of length $2^m$, we have
\begin{equation}\label{eq:G_23}
\Gbu_2 ^{(m)} =
\begin{bmatrix}
{\bf 1} \\
{\xbu}_0 \\
%x_1 \\
\vdots \\
{\xbu}_{m-1} \\
{\xbu}_{m-1} \xbu_0 \\
%{\xbu}_{m-1} x_1 \\
\vdots \\
\xbu_{m-1} \xbu_{m-2} \\
\end{bmatrix}, \quad
\Gbu_3 ^{(m)}  =
\begin{bmatrix}
{\bf 1} \\
{\xbu}_0 \\
%x_1 \\
\vdots \\
{\xbu}_{m-1} \\
{\xbu}_{m-1} \xbu_0 \\
%{\xbu}_{m-1} x_1 \\
\vdots \\
\xbu_{m-1} \xbu_{m-2} \\
\xbu_{m-2} \xbu_0 \\
\vdots \\
\xbu_{m-2} \xbu_{m-3} \\
\xbu_{m-1} \xbu_{m-2} \xbu_0\\
\vdots \\
\xbu_{m-1} \xbu_{m-2} \xbu_{m-3} \\
\end{bmatrix}.
\end{equation}
Note that $\Gbu_2 ^{(m)}$ and $\Gbu_3 ^{(m)}$ in (\ref{eq:G_23})
have the different orders of rows with those generated by the recursions of (\ref{eq:G_r_org}) or (\ref{eq:G_r}).
In this paper, we use $\Gbu_2 ^{(m)}$ and $\Gbu_3 ^{(m)}$ in (\ref{eq:G_23}).
%which provide lower PAPRs for more users.

\vspace{0.1in}
\begin{exa}
Let $m=3$. Then, the generator matrices of ${\cal B}_2$ and ${\cal B}_3$ are $6 \times 8$ and
$8 \times 8$ matrices, respectively.
\begin{equation*}
\Gbu_2 ^{(3)}=
\begin{bmatrix}
1111 & 1111  \\
0101 & 0101  \\
0011 & 0011  \\
0000 & 1111  \\
0000 & 0101  \\
0000 & 0011  \\
\end{bmatrix} =
\begin{bmatrix}
{\bf 1} \\
{\xbu}_0 \\
{\xbu}_1 \\
{\xbu}_2 \\
{\xbu}_{2} \xbu_0 \\
\xbu_2 \xbu_1 \\
\end{bmatrix}, \quad
\Gbu_3  ^{(3)} =
\begin{bmatrix}
1111 & 1111  \\
0101 & 0101  \\
0011 & 0011  \\
0000 & 1111  \\
0000 & 0101  \\
0000 & 0011  \\
0001 & 0001  \\
0000 & 0001  \\
\end{bmatrix} =
\begin{bmatrix}
{\bf 1} \\
{\xbu}_0 \\
{\xbu}_1 \\
{\xbu}_2 \\
{\xbu}_{2} \xbu_0 \\
\xbu_2 \xbu_1 \\
\xbu_1 \xbu_0 \\
\xbu_2 \xbu_1 \xbu_0 \\
\end{bmatrix}.
\end{equation*}
\end{exa}
\vspace{0.1in}

%If $N > 1$,
%the encoding process is applied to $\abu_n$ at each $n$th spreading for $0 \leq n \leq N-1$, i.e.,
%$\bbu_n = \abu_n \cdot \Gbu_r ^{(m)}$.

We briefly introduce some decoding techniques for Reed-Muller subcodes.
The first-order Reed-Muller code can be decoded by
the \emph{Fast Hadamard Transform (FHT)} technique described in~\cite{Wicker:ECC}.
In general, the $r$th-order Reed-Muller code
is decoded by the \emph{Reed decoding algorithm}~\cite{Reed:RM}.
In particular, if we consider ${\cal B}_2 ^{(m)}$ or ${\cal B}_3 ^{(m)}$ as a supercode
of the union of cosets of ${\rm R}(1, m)$, then
we can accomplish the soft decision decoding by removing each possible coset representative
from the received codeword and then applying the FHT~\cite{ConSloane:soft}\cite{DavJed:GDJ}.

For the encoding and decoding of Golay complementary sequences, see~\cite{DavJed:GDJ}.

\subsection{PAPR of coded MC-CDMA signals with $N > 1$}
In what follows, we discuss the PAPR of coded MC-CDMA signals in a general case of $N>1$.
We restrict our attention to a Walsh-Hadamard spread MC-CDMA employing ${\cal B}_2 ^{(m)}$ or ${\cal B}_3 ^{(m)}$
%as its coding scheme, % described in Section IV,
that provides the acceptable code rate as well as the low PAPR for the coded MC-CDMA signals.
We show that the maximum PAPR depends on the actual number of users supported by the MC-CDMA.
%where ${\cal B}_2 ^{(m)}$ and ${\cal B}_3 ^{(m)}$ are employed as a coding scheme.

\vspace{0.1in}

\begin{thr}\label{th:user}
Assume that ${\cal B}_2 ^{(m)}$ is employed
in a Walsh-Hadamard spread MC-CDMA system in Figure~\ref{fig:coded_sys}, where $K=L=2^m$.
The maximum PAPR of the coded MC-CDMA signal $s(t)$ is then determined by % the number of actual users $w$, i.e.,
\begin{equation}\label{eq:papr2_w}
{\rm PAPR}(s(t))\leq \left\{ \begin{array}{ll} N, \quad & \mbox{ if } 1 \leq w \leq m+1,\\
2N, \quad & \mbox{ if } m+2 \leq w \leq 2m. \end{array} \right.
\end{equation}
Similarly, if the system employs ${\cal B}_3 ^{(m)}$,
then the maximum PAPR of $s(t)$ is % given by
\begin{equation}\label{eq:papr3_w}
{\rm PAPR}(s(t))\leq \left\{ \begin{array}{ll} N, \quad & \mbox{ if } 1 \leq w \leq m+1,\\
2N, \quad & \mbox{ if } m+2 \leq w \leq 2m, \\
4N, \quad & \mbox{ if } 2m+1 \leq w \leq 4m-4. \end{array} \right.
\end{equation}
\end{thr}
\vspace{0.1in}

\noindent \emph{Proof.}
In Theorem~\ref{th:papr_s_n}, it is easy to see that
$P=\max_{0 \leq n \leq N-1} {\rm PAPR}(s_n(t)) = \max_{\dbu_0} {\rm PAPR}(s_0(t))$,
where $s_0(t)$ and $s_n(t)$ are given in (\ref{eq:s_t_0}) and (\ref{eq:s_n_2}), respectively.
%when a coding scheme is repeatedly applied for every $n$, $0 \leq n \leq N-1$.
%note that $P=\max_{0 \leq n \leq N-1} {\rm PAPR}(s_n(t))$ is equal to the maximum of $ {\rm PAPR}(s_0(t))$
%for all possible $\dbu_0$'s,
Therefore, $P=1, 2$, and $4$ when ${\cal B}_1 ^{(m)}$, ${\cal B}_2 ^{(m)}$, and ${\cal B}_3 ^{(m)}$
are employed as the coding scheme, respectively.
In $\Gbu_2 ^{(m)}$ of (\ref{eq:G_23}),
if $w \leq m+1$, the first $(m+1)$ rows participate in the encoding process, while
the other rows are ignored by zero tailing.
Since a linear combination of the first $(m+1)$ rows generates a codeword of ${\cal B}_1 ^{(m)}$, % (or ${\cal B}_1 ^{(m)}$ in this context),
it is obvious that if $w \leq m+1$, then ${\rm PAPR}(s(t))\leq N$ from Theorems~\ref{th:papr_s_n} and \ref{th:papr_1}.
If $m+2 \leq w \leq 2m$, on the other hand, ${\rm PAPR}(s(t))\leq 2N$ from Construction~\ref{cst:B2} and Theorem~\ref{th:papr_s_n}.
Therefore, (\ref{eq:papr2_w}) is true for ${\cal B}_2 ^{(m)}$.
Similar to this approach, (\ref{eq:papr3_w}) is also true
for ${\cal B}_3 ^{(m)}$ from the generator matrix $\Gbu_3 ^{(m)}$ of (\ref{eq:G_23}) and Theorem~\ref{th:papr_s_n}.
\qed
\vspace{0.1in}

In general, if ${\cal B}_r ^{(m)}$ is employed as the coding scheme,
the Walsh-Hadamard spread MC-CDMA signals have the PAPR of at most $N \cdot 2^{r-1}$ from Construction~\ref{cst:gen_code} and Theorem~\ref{th:papr_s_n}.
However, Theorem~\ref{th:user} is not true for the MC-CDMA signals % encoded by ${\cal B}_r ^{(m)}$
if the generator matrix $\Gbu_r ^{(m)}$ is
recursively constructed by (\ref{eq:G_r_org}) or (\ref{eq:G_r}).
We need to reorder the rows of $\Gbu_r ^{(m)}$ % according to the Boolean expression as in (\ref{eq:G_23})
to achieve the maximum PAPR
depending on the number of actual users as in Theorem~\ref{th:user}. % for ${\cal B}_r ^{(m)}$.

\vspace{0.1in}
\begin{rem}
In Section IV-B, we developed various Reed-Muller subcodes of length $2^m$ to control the peak power of MC-CDMA signals with
$2^m$ subcarriers in a systematic way. %, where $N=1$.
In fact,
we may employ the coding scheme for a codeword of length $2^m N $, where $N = 2^{h}$,
which encodes $\abu = (\abu_0 \ | \ \abu_1 \ | \ \cdots \ | \ \abu_{N-1})$, a concatenation of $N$ uncoded data block.
Then, the codeword covers the entire $2^m N $ subcarriers to control the peak power
and to ultimately reduce the maximum PAPR of the coded MC-CDMA signal.
In this case, however, the code rate may be dramatically reduced
for such a long codeword
because the coding scheme is a subcode of the Reed-Muller code.
This also enlightens a connection between the code rates and the maximum PAPR of coded MC-CDMA signals.
\end{rem}
\vspace{0.1in}

%\vspace{0.1in}
\begin{rem}
%Supporting a single user by % a coded MC-CDMA system
Virtually treating a single user's data as
multiple users' one, % in a system,
a coded MC-CDMA system can be considered as an equivalent \emph{spread OFDM}~\cite{Deb:SOFDM},
where the single user's data is spread across a set of subcarriers to enjoy frequency diversity.
By applying the coding schemes introduced in this section,
the spread OFDM additionally has the benefits of low PAPR and good error correction capability.
\end{rem}
%\vspace{0.1in}

\section{Simulation Results and Discussions}

This section provides simulation results to confirm our theoretical analysis
and presents some discussions on statistical results of PAPR of MC-CDMA signals.
The PAPR properties of a Reed-Muller coded and Walsh-Hadamard spread MC-CDMA system are compared to
those of a pair of uncoded systems.
%described in~\cite{OchImai:OFDM-CDMA}.
In the uncoded systems, the one employs Walsh-Hadamard (WH) spreading sequences, while the other uses
Golay complementary (GC) spreading sequences
each of which forms a row of a recursively constructed
Golay complementary spreading matrix~\cite{OchImai:OFDM-CDMA}.
In our coded MC-CDMA, we employ ${\cal B}_3 ^{(m)}$ as the coding scheme
%with $W=4(m-1)$ and $K=2^m$,
which we believe is a good coding solution providing the acceptable code rates, the moderate complexity, and the low PAPR
for the coded MC-CDMA signals.

For a fair comparison, we assume that all the MC-CDMA systems transmit the same number of information bits in an OFDM symbol
from $w$ active users.
If the uncoded systems transmit $Nw$ information bits in an OFDM symbol,
our coded system of code rate $R$ then needs to transmit $\frac{Nw}{R}$ \emph{coded} bits
for the transmission of $Nw$ information bits.
Therefore, while the uncoded ones have $NL$ subcarriers,
the coded system needs to use $\frac{NL}{R}$ subcarriers in an OFDM symbol,
where $L$ is a spreading factor used in the uncoded systems.
In the following, our simulations employ ${\cal B}_3 ^{(m)}$ of code rate $R=0.5$ or $1$,
where the coded MC-CDMA uses $2NL$ or $NL$ subcarriers in an OFDM symbol
%to transmit $Nw$ information bits from $w$ users
by employing the
spreading sequences of length $2L$ or $L$.

%When we measure the PAPR of each MC-CDMA signal,
In our simulations,
we measure the discrete-time PAPR~\cite{Tell:comp} of each MC-CDMA signal
from the IDFT (Inverse discrete Fourier transform) of
the oversampling factor $8$.
Also, we statistically measure the PAPR over $N_s=5 \times 10^5$ OFDM symbols
for $NwN_s$ randomly generated information bits.

\subsection{Code rate $R=0.5$}

%Since the spreading sequences of length $L=2^m$ are assigned \emph{on demand}, $K=W=w$ in the uncoded systems
%supporting $w$ active users.
%In our coded MC-CDMA, on the other hand, we employ ${\cal B}_3 ^{(m)}$ as a coding scheme
%with $W=4(m-1)$ and $K=2^m$,
%which we believe is a good coding solution providing acceptable code rates and low PAPRs
%for the coded MC-CDMA signals.
%As a fully-loaded system, the coded MC-CDMA has $K=L=2^m$.

\begin{figure}[t!]
\centering
\includegraphics[width=1.0\textwidth, angle=0]{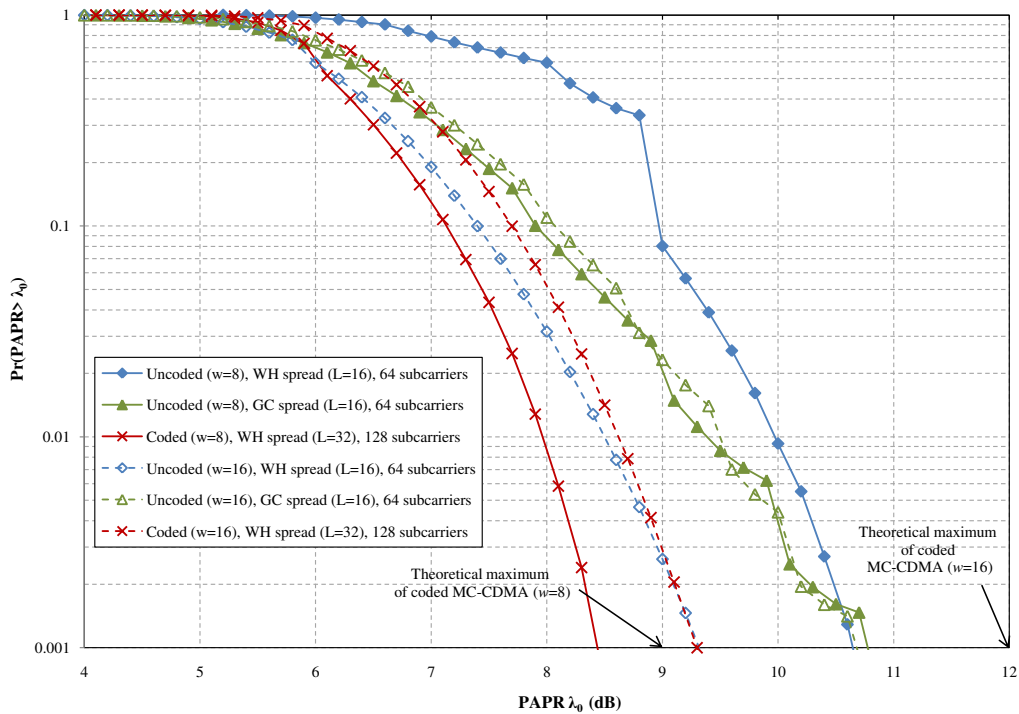}
\caption{PAPR performance of MC-CDMA systems.
%In uncoded systems, the spreading factor and the number of subcarriers are $16$ and $64$, respectively.
%The coded system has a spreading factor $32$ and the number of subcarriers $128$.
All the MC-CDMA systems transmit $Nw=32$ or $64$ information bits in an OFDM symbol.
The code rate of the coded MC-CDMA is $0.5$.}
\label{fig:CCPD12}
\end{figure}

\begin{figure}[t!]
\centering
\includegraphics[width=1.0\textwidth, angle=0]{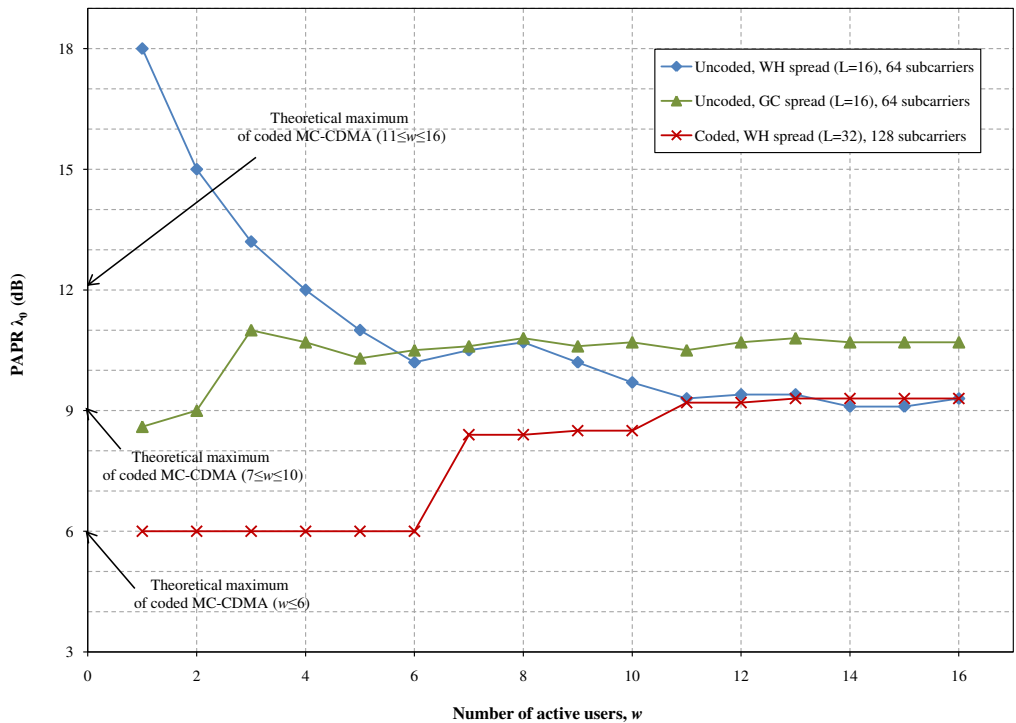}
\caption{Relationship between PAPR $\lambda_0$ and active number of users for MC-CDMA systems
where ${\rm Pr( PAPR} > \lambda_0) = 10^{-3}$.
%In uncoded systems, the spreading factor and the number of subcarriers are $16$ and $64$, respectively.
%The coded system has a spreading factor $32$ and the number of subcarriers $128$.
The code rate of the coded MC-CDMA is $0.5$.}
\label{fig:PAPR12}
\end{figure}

In Figures~\ref{fig:CCPD12} and \ref{fig:PAPR12}, $w$ users access to each MC-CDMA system
to transmit $N=4$ information bits per each user in an OFDM symbol.
%In uncoded MC-CDMA systems,
%the spreading sequences are assigned \emph{on demand}, so $K=W=w$.
The uncoded MC-CDMA systems use the spreading factor $L=16$ and $NL=64$ subcarriers, where
maximum $16$ users are supported, i.e., $1 \leq w \leq 16$.
As the coded MC-CDMA system also needs to support up to $16$ users, we choose a $(32, 16)$ code ${\cal B}_3 ^{(5)}$
as its coding scheme,
where $W=16$, $K=32$, and $R=0.5$.
Thus, our coded MC-CDMA uses the spreading factor $2L=32$ and $2NL=128$ subcarriers
to transmit $Nw$ information bits in an OFDM symbol from the $w$ active users.
In the coded MC-CDMA system, each $32$-bit codeword is fully-loaded to all the available $32$ Walsh-Hadamard spreading sequences
regardless of $w$.
On the other hand,
the uncoded MC-CDMA systems assign the $w$ spreading sequences to $w$ users \emph{on demand},
so they are fully-loaded only if $w=16$.

Figure~\ref{fig:CCPD12} shows the complementary cumulative distribution functions (CCDF) of ${\rm Pr}({\rm PAPR} > \lambda_0)$ of
each MC-CDMA signal for $w=8$ and $16$.
It reveals that the coded MC-CDMA is superior to the others when the number of active users is small.
Precisely, if $w=8$, it reduces the PAPR $\lambda_0$ achieving ${\rm Pr ( PAPR} > \lambda_0) = 10^{-3}$
by more than $2$ dB, compared to the uncoded systems.
Moreover, Theorem~\ref{th:user} ensures that there exists no coded MC-CDMA signal with PAPR $> 9$ dB for $w=8$,
which implies that the coded MC-CDMA also outperforms the uncoded ones in theoretical aspects.
If $w=16$, the coded MC-CDMA has almost the same PAPR $\lambda_0$ as the uncoded Walsh-Hadamard spread MC-CDMA
for achieving ${\rm Pr ( PAPR} > \lambda_0) = 10^{-3}$.
Even in this case, it is theoretically guaranteed that no coded MC-CDMA signal has ${\rm PAPR} > 12$ dB, % in the coded MC-CDMA,
which may not be true in the uncoded systems.
Figure~\ref{fig:CCPD12} also shows that most of the coded MC-CDMA signals in the statistical experiments
have much smaller PAPR than
the theoretical maximum predicted by Theorem~\ref{th:user}.

Figure~\ref{fig:PAPR12} displays the PAPR $\lambda_0$ of each MC-CDMA achieving ${\rm Pr ( PAPR} > \lambda_0) = 10^{-3}$
according to the number of active users $w, \ 1 \leq w \leq 16$.
It is well known~\cite{OchImai:OFDM-CDMA} that the uncoded Walsh-Hadamard (WH) spread MC-CDMA shows the high PAPR
when the number of active users is small.
The PAPR then decreases as the number of users increases.
On the other hand, the uncoded Golay complementary (GC) spread MC-CDMA has the low PAPR for the small number of users.
However, the PAPR gets higher than that of the uncoded Walsh-Hadamard spread MC-CDMA
as the number of users increases.
Figure~\ref{fig:PAPR12} shows that
the coded MC-CDMA is a good alternative to the two uncoded systems
by providing the smallest PAPR $\lambda_0$ for almost all user numbers.
Moreover, Theorem~\ref{th:user} assures that the maximum PAPR of the coded system is theoretically limited to $6$ dB for $1 \leq w \leq 6$,
$9 $ dB for $ 7 \leq w \leq 10$, and $12 $ dB for $ 11 \leq w \leq 16$, respectively,
where Figure~\ref{fig:PAPR12} provides the numerical evidences.
Therefore, it is theoretically guaranteed % by Theorem~\ref{th:user}
that there exists no coded MC-CDMA signal with the PAPR higher than the maximum values for each user,
which is not generally true in the uncoded systems.
The theoretical and statistical results show that
the coded MC-CDMA dramatically reduces its PAPR % uncoded ones in theoretical and statistical aspects
for the small number of users, % are supported by the system,
which effectively solves the high PAPR problem in the uncoded MC-CDMA.
Figure~\ref{fig:PAPR12} also shows that if the number of active users is large ($w \geq 11$),
the statistical PAPR $\lambda_0$ is much smaller than the theoretical maximum predicted by Theorem~\ref{th:user}.
As a result, we claim that the coded MC-CDMA provides the best statistical and theoretical solution
for PAPR reduction for any number of users.

\subsection{Code rate $R=1.0$}

\begin{figure}[t!]
\centering
\includegraphics[width=1.0\textwidth, angle=0]{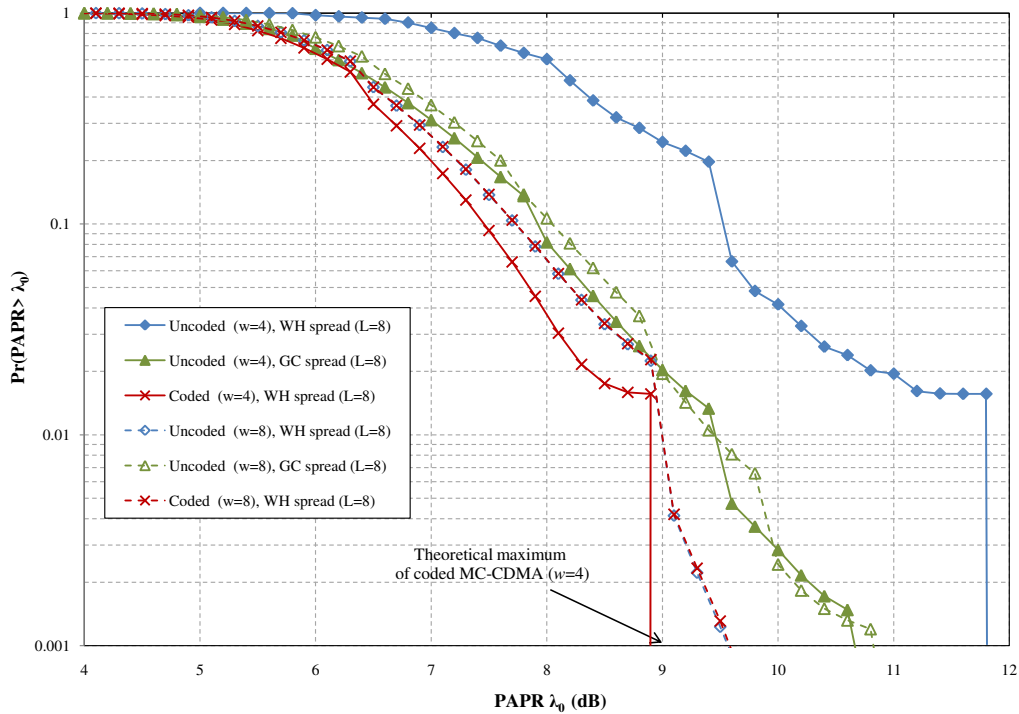}
\caption{PAPR performance of MC-CDMA systems.
All the MC-CDMA systems transmit $Nw=32$ or $64$ information bits in an OFDM symbol employing
the spreading factor $8$ and $64$ subcarriers.
The code rate of the coded MC-CDMA is $1.0$.}
\label{fig:CCPD1}
\end{figure}

\begin{figure}[t!]
\centering
\includegraphics[width=1.0\textwidth, angle=0]{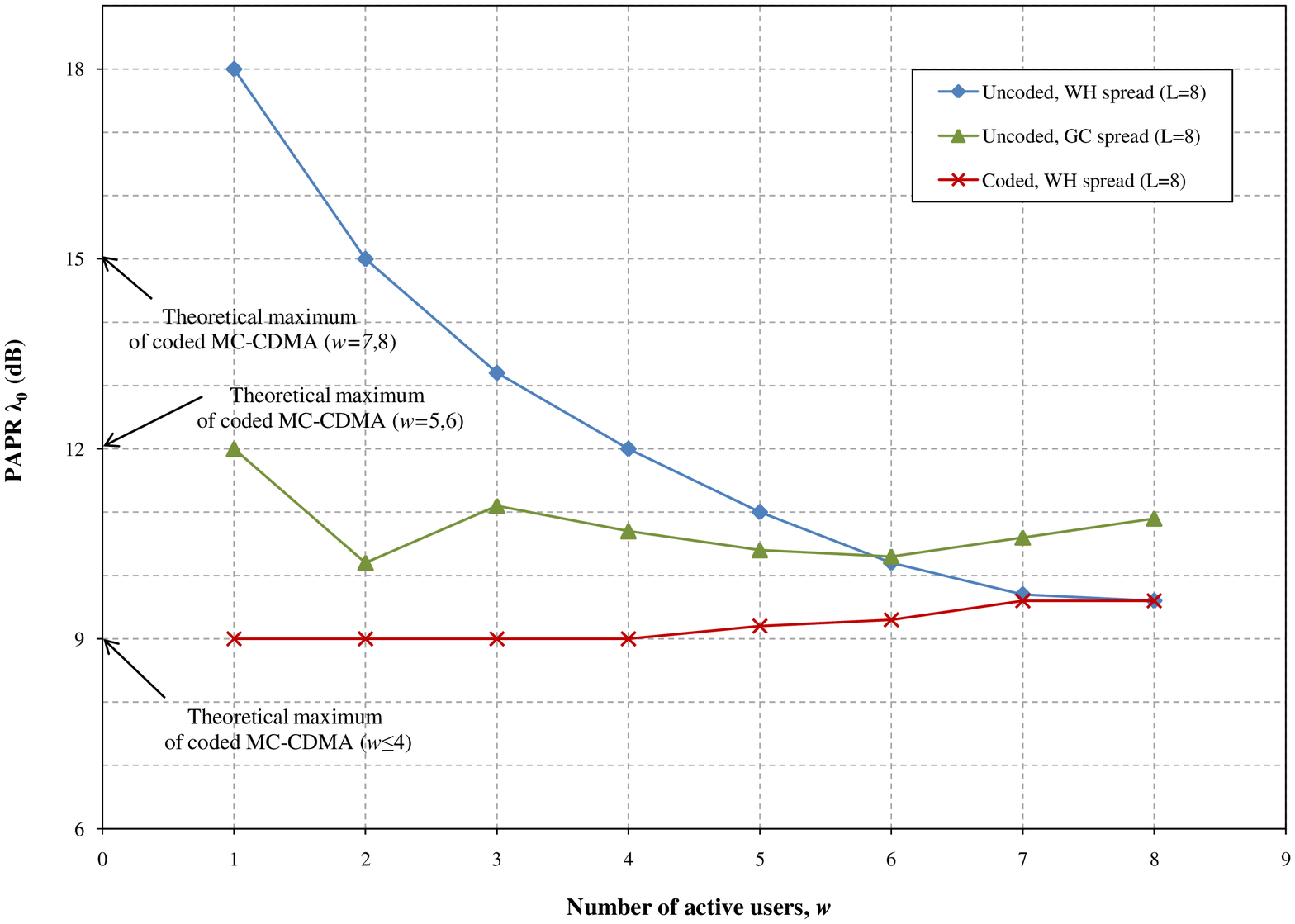}
\caption{Relationship between PAPR $\lambda_0$ and active number of users for MC-CDMA systems
where ${\rm Pr( PAPR} > \lambda_0) = 10^{-3}$.
All the MC-CDMA systems use
the spreading factor $8$ and $64$ subcarriers. The code rate of the coded MC-CDMA is $1.0$.}
\label{fig:PAPR1}
\end{figure}

In Figures~\ref{fig:CCPD1} and \ref{fig:PAPR1},
the MC-CDMA systems support $w$ users where each user transmits $N=8$ information bits
in an OFDM symbol.
The uncoded MC-CDMA systems use the spreading factor $L=8$ and $NL=64$ subcarriers
to transmit $Nw$ information bits in an OFDM symbol.
To support up to $8$ users,
the coded MC-CDMA system employs ${\cal B}_3 ^{(3)}$, a $(8, 8)$ code with $K=W=8$ and $R=1$.
In this case, ${\cal B}_3 ^{(3)}$ is used as a mapping scheme
mentioned in Remark~\ref{rem:RM_map}.
%where the theoretical maximum of PAPR is $4$ for a single spreading process.
In our MC-CDMA, each $8$-bit uncoded data $\abu_n$, $0 \leq n \leq 7$, is \emph{transformed}
by the Reed-Muller mapping scheme for PAPR reduction.
Thus, it uses the spreading factor $8$ and $64$ subcarriers
to transmit $Nw$ data bits in an OFDM symbol, which is the same as the uncoded systems.
However, note that % our MC-CDMA has a zero-tail process, and
each $8$-bit codeword of our MC-CDMA is fully-loaded to all the available
spreading sequences of length $8$ regardless of $w$.
%where all the available $8$ spreading sequences are assigned to an $8$-bit codeword.

Figure~\ref{fig:CCPD1} shows the results of ${\rm Pr}({\rm PAPR} > \lambda_0)$ of each MC-CDMA for $w=4$ and $8$.
We observed that if $w=4$, our Reed-Muller mapped MC-CDMA system reduces the PAPR $\lambda_0$ by about $3$ dB
to achieve ${\rm Pr}({\rm PAPR} > \lambda_0)=10^{-3}$,
compared to the uncoded Walsh-Hadamard (WH) spread MC-CDMA employing the same spreading factor and the same number of subcarriers.
Moreover, Theorem~\ref{th:user} ensures that our system has no signal with PAPR $> 9$ dB for $w=4$.
Note that Theorem~\ref{th:user} determines the maximum PAPR of $9$ dB (if $w \leq 4$),
$12$ dB (if $5 \leq w \leq 6$), and $15$ dB (if $7 \leq w \leq 8$), respectively.
Thus, even if the statistical PAPR property of our MC-CDMA is almost identical to that of the uncoded Walsh-Hadamard spread MC-CDMA
for $w=8$,
our system has no probability of signals with the PAPR higher than $15$ dB, % the maximum for the corresponding number of users,
which is however unclear in the uncoded systems.

Similar to Figure~\ref{fig:PAPR12}, Figure~\ref{fig:PAPR1} displays the PAPR $\lambda_0$ of each MC-CDMA
achieving ${\rm Pr ( PAPR} > \lambda_0) = 10^{-3}$
according to the number of users $w, 1 \leq w \leq 8$.
%It is well known that the uncoded Walsh-Hadamard spread MC-CDMA shows high PAPR
%when the number of active users is small.
%The PAPR then decreases as the number of active users increases.
%On the other hand, the uncoded Golay complementary spread MC-CDMA has smaller PAPR for a small number of users.
%However, the PAPR gets higher than that of uncoded Walsh-Hadamard MC-CDMA
%as the number of users increases.
It shows that
our MC-CDMA % is a good compromise between the two uncoded systems
provides the smallest PAPR $\lambda_0$ for almost all user numbers.
Also, it numerically confirms that the theoretical maximums of PAPR in Theorem~\ref{th:user} hold
for each user number.
%Theorem~\ref{th:user} ensures that the maximum PAPR of the coded system is $6$ dB for $1 \leq w \leq 6$,
%$9 $ dB for $ 7 \leq w \leq 10$, and $12 $ dB for $ 11 \leq w \leq 16$, respectively,
%where Figure provides the numerical evidences.
Figure~\ref{fig:PAPR1} shows that if the number of active users is large,
the statistical PAPR $\lambda_0$ is much smaller than the theoretical maximum predicted by Theorem~\ref{th:user}.
For the small number of users, on the other hand, % it turns out that % we see that
the coded MC-CDMA solves the high PAPR problem of the uncoded MC-CDMA
by dramatically reducing its PAPR. % uncoded ones in theoretical and statistical aspects % supported by the system,
Along with Figure~\ref{fig:PAPR12},
% we claim that
the coded MC-CDMA can be the best statistical and theoretical solution for PAPR reduction
for any number of active users regardless of its code rate.
A drawback of the Reed-Muller mapped MC-CDMA is that it could be employed only for a small spreading factor
due to the demapping complexity at the receiver.

\section{Conclusion}

This paper has presented a coded MC-CDMA system where
the information data is encoded by a Reed-Muller subcode
for the sake of PAPR reduction.
In the system, the codeword is then fully-loaded to Walsh-Hadamard spreading sequences,
where the spreading and the despreading processes are efficiently implemented by the Walsh-Hadamard transform (WHT).
%and the spread data is finally allocated to each subcarrier after interleaving.
We have established the polynomial representation of
a coded MC-CDMA signal for
theoretical analysis of the PAPR.
We have then developed a recursive construction of the Reed-Muller subcodes which
provide the transmitted MC-CDMA signals with the bounded PAPR
as well as the error correction capability.
We have also investigated a theoretical connection between the code rates and the maximum PAPR
in the coded MC-CDMA.
Simulation results showed that
the PAPR of the coded MC-CDMA signal is not only theoretically bounded, but also statistically reduced
by the Reed-Muller coding schemes.
In particular, it turned out that the coded MC-CDMA could
solve the PAPR problem of uncoded MC-CDMA
by dramatically reducing its PAPR
for the small number of users.
Finally, the theoretical and statistical studies exhibited that
the Reed-Muller subcodes are effective coding schemes for peak power control in MC-CDMA
with small and moderate numbers of users, subcarriers, and spreading factors.
We believe this work gives us theoretical insights
for PAPR reduction of MC-CDMA and S-OFDM
by means of an error correction coding.


%\section*{Acknowledgment}




%\bibliographystyle{IEEEtran}

\begin{thebibliography}{1}



\bibitem{802.11} IEEE Standard 802.11-2007, IEEE Standard for Information Technology -
Local and Metropolitan Area Networks - Specific Requirements,
Part 11 - Wireless LAN Medium Access Control (MAC) and Physical Layer (PHY) Specifications.

\bibitem{WiMAX} IEEE802.16e-2005, IEEE Standard for Local and Metropolitan Area Networks,
Part 16 - Air Interface for Fixed and Mobile Broadband Wireless Access Systems,
Amendment 2: Physical and Medium Access Control Layers for Combined Fixed and Mobile
Operation in Licensed Bands.

\bibitem{LTE:36-211} 3GPP TS 36.211, v.~8.3.0,
Technical Specification Group Radio Access Network;
Evolved Universal Terrestrial Radio Access (E-UTRA);
Physical Channels and Modulation
(Release 8).



\bibitem{Yee:MC-CDMA}
N.~Yee, J.~P.~Linnartz, and G.~Fettweis,
``Multi-carrier CDMA in indoor wireless radio networks,''
\emph{in Proc. of IEEE PIMRC}, pp.~109-113, Sep. 1993.

\bibitem{Fazel:MC-CDMA}
K.~Fazel, S.~Kaiser, and M.~Schnell,
``A flexible and high performance celluar mobile communications systems based on
orthogonal multicarrer SSMA,''
\emph{Wireless Personal Communications}, vol.~2, pp.~121-144, 1995.

\bibitem{Hara:MC-CDMA}
S.~Hara and R.~Prasad,
``Overview of multicarrier-CDMA,''
\emph{IEEE Commun. Mag.}, vol.~35, no.~12, pp.~126-133, Dec. 1997.


\bibitem{Pat:GRM} K.~G.~Paterson,
``Generalized Reed-Muller codes and power control in OFDM modulation,''
\emph{IEEE Trans. Inform. Theory}, vol.~46, no.~1, pp.~104-120, Jan. 2000.




\bibitem{Jones:block}
A.~E.~Jones, T.~A.~Wilkinson, and S.~K.~Barton,
``Block coding scheme for reduction of peak to mean envelope power ratio of
multicarrier communication schemes,''
\emph{Electron. Lett.}, vol.~30, pp.~2098-2099, 1994.



\bibitem{Nee:OFDM}
R.~D.~J.~van Nee,
``OFDM codes for peak-to-average power reduction and error correction,''
\emph{in Proc. of IEEE GLOBECOM}, London, U.K., pp.~740-744, 1996.



\bibitem{Golay}
M.~J.~E.~Golay, ``Complementary series,''
\emph{IRE Trans. Inform. Theory}, vol.~IT-7, pp.~82-87, 1961.



\bibitem{DavJed:GDJ}
J.~A.~Davis and J.~Jedwab,
``Peak-to-mean power control for OFDM, Golay complementary sequences, and Reed-Muller codes,''
\emph{IEEE Trans. Inform. Theory}, vol.~45, no.~7, pp.~2397-2417, Nov. 1999.



\bibitem{Pat:Codes} K.~G.~Paterson,
``On codes with low peak-to-average power ratio for multicode CDMA,''
\emph{IEEE Trans. Inform. Theory}, vol.~50, no.~3, pp.~550-559, Mar. 2004.


\bibitem{Pat:alg} K.~G.~Paterson,
``Sequences for OFDM and multi-code CDMA: Two problems in algebraic coding theory,''
HPL-2001-146, Hewlett-Packard Laboratories, 2001.




\bibitem{Han:overview}
S.~H.~Han and J.~H.~Lee,
``An overview of peak-to-average power ratio reduction techniques for multicarrier transmission,''
\emph{IEEE Wireless Communications}, pp. 56-65, 2005.



\bibitem{OchImai:OFDM-CDMA}
H.~Ochiai and H.~Imai,
``OFDM-CDMA with peak power reduction based on the spreading sequences,''
\emph{IEEE Int. Conf. on Commun. (ICC)}, vol.~3, pp. 1299-1303, 1998.



\bibitem{Pop:spreading}
B.~M.~Popovi\'c,
``Spreading sequences for multicarrier CDMA systems,''
\emph{IEEE Trans. Commun.}, vol.~47, no.~6, pp.~918-926, 1999.


\bibitem{Nob:SS}
S.~Nobilet, J.~F.~Hlard, and D.~Mottier,
``Spreading sequences for uplink and downlink MC-CDMA: PAPR and MAI minimaization,''
\emph{European Trans. Telecommun.}, vol.~13, pp.~465-473, Sept./Oct. 2002.


\bibitem{ChoiHanzo:CF}
B.~-J.~Choi, E.-L.~Kuan, and L.~Hanzo,
``Crest-factor study of MC-CDMA and OFDM'',
\emph{IEEE VTC'99(Fall)}, pp.~233-237, 1999.


\bibitem{ChoiHanzo:CFCS}
B.~-J.~Choi and L.~Hanzo,
``Crest factors of complementary-sequence-based multicode MC-CDMA sginals,''
\emph{IEEE Trans. on Wireless Commun.}, vol.~2, no.~6, pp.~1114-1119, Nov. 2003.


\bibitem{Pog:SS}
E.~Pogossova, K.~Egiazarian, and J.~Astola,
``Spreading sequences for downlink MC-CDMA transmission,''
\emph{IEEE VTC2004(Fall)}, vol.~7, pp.~4859-4863, 2004.

\bibitem{Yang:alloc}
L.~Yang and E.~Alsusa,
``Dynamic code-allocation based PAPR reduction technique for MC-CDMA systems,''
\emph{in Proc. of WCNC 2007}, pp.~628-633, 2007.

\bibitem{Shi:sym}
Q.~Shi and Q.~T.~Zhang,
``Symmetry-embedded spreading sequences for multicarrier CDMA,''
\emph{IEEE Trans. on Wireless Commun.}, vol.~6, no.~10, pp.~3534-3539, Oct. 2007.



\bibitem{WieWu:High}
D.~A.~Wiegandt, Z.~Wu, and C.~R.~Nassar,
``High-throughput, high-performance OFDM via pseudo-orthogonal carrier interferometry
spreading codes,''
\emph{IEEE Trans. Commun.}, vol.~51, no.~7, pp.~1123-1134, July 2003.


\bibitem{Kaiser:SOFDM}
S.~Kaiser,
``On the performance of different detection  techniques for OFDM-CDMA in fading channels,''
\emph{in Proc. of IEEE GLOBECOM}, pp.~2059-2063, 1995.



\bibitem{Deb:SOFDM}
D.~Merouane, P.~Loubaton, and M.~de Courville,
``Spread OFDM performance with MMSE equalization,''
\emph{in Proc. of IEEE Conf. Acoustics, Speech, Signal Processing}, pp.~2385-2388, 2001.

\bibitem{Al:CSOFDM}
M.~Al-Mahmoud,
``Perfomance evaluation of code-spread OFDM,''
\emph{46th Annual Allerton Conf.}, Allerton House, University of Illinois-Urbana Champaigne, Illinois, USA,
pp.~274-278, Sept. 2008.



\bibitem{Hanzo:book}
L.~Hanzo, M.~M\"unster, B.~-J.~Choi, and T.~Keller,
\emph{OFDM and MC-CDMA for Broadband Multi-user Communications, WLANs and Broadcasting},
John Wiley \& Sons, 2003.


\bibitem{Ye:FEC}
Z.~Ye, G.~J.~Saulnier, D.~Lee, and M.~J.~Medley,
``FEC coding with a rate adaptive spread spectrum OFDM system,''
\emph{Communication Theory Mini-Conference}, Vancouver, BC, Canada,
pp.~67-71, June 1999.

\bibitem{Parker:close}
M.~G.~Parker,
``Close encounters with Boolean functions of three different kinds,''
\emph{2nd International Castle Meeting on Coding Theory and Applications}, 
Valladolid, September 2008.
Also available at \emph{Lecture Notes in Computer Science (LNCS)} vol.~5228, pp.~137-153, 2008.


\bibitem{Mac:ECC}
F.~J.~MacWillams and N.~J.~A.~Sloane,
\emph{The Theory of Error Correcting Codes},
Amsterdam, The Netherlands: North Holland, 1986.


\bibitem{SebYam:Had}
J.~Seberry and M.~Yamada,
``Hadamard matrices, sequences, and block designs,''
\emph{Contemporary Design Theory: A Collection of Surveys},
J.~H.~Dinitz and D.~R.~Stinson, Eds. John Wiley \& Sons, Inc. 1992.


\bibitem{Lit:ppc}
S.~Litsyn,
\emph{Peak Power Control in Multicarrier Communications}, Cambridge University Press, 2007.



\bibitem{ParkKen:GCS}
M.~G.~Parker, K.~G.~Paterson, and C.~Tellambura,
``Golay complementary sequences,''
\emph{Wiley Encyclopedia of Telecommunications}, Edited by J.~G.~Proakis, Wiley Interscience, 2002.



\bibitem{ParkTell:GDJ}
M.~G.~Parker and C.~Tellambura,
``Golay-Davis-Jedwab complementary sequences and Rudin-Shapiro constructions,''
\emph{manuscript.} 2001. Available in ``http://www.ii.uib.no/$\sim$matthew/ConstaBent2.pdf''.


%\bibitem{ParkTell:WCC}
%M.~G.~Parker and C.~Tellambura,
%``Generalised Rudin-Shapiro constructions,''
%\emph{in Proc. of Workshop on Coding and Cryptography (WCC)}, Paris, France, 2001.
%Available in ``www.ii.uib.no/$\sim$matthew/RudinShap2.ps''.



\bibitem{Wicker:ECC}
S.~B.~Wicker, \emph{Error Control Systems for Digital Communication and Storage,}
Englewood Cliffs: Prentice-Hall, 1995.


\bibitem{Reed:RM}
I.~S.~Reed,
``A class of multiple-error correcting codes and a decoding scheme,''
\emph{IEEE Trans. Inform. Theory}, vol.~4, pp.~38-49, Sep. 1954.

\bibitem{ConSloane:soft}
J.~H.~Conway and N.~J.~A.~Sloane,
``Soft decoding techniques for codes and lattices, including the Golay codea and the Leech lattice,''
\emph{IEEE Trans. Inform. Theory}, vol.~IT-32, no.~1, pp.~41-50, Jan. 1986.


\bibitem{Tell:comp}
C.~Tellambura,
``Computation of the continuous-time PAR of an OFDM signal with BPSK subcarriers,''
\emph{IEEE Commun. Letters}, vol.~5, no.~5, pp.~185-187, May 2001.


%\bibitem{BorFug:comp}
%P.~B.~Borwein and R.~A.~Ferguson,
%``A complete description of Golay pairs for lengths up to $100$,''
%\emph{Mathematics of Computation}, vol.~73, no.~246, pp.~967-985, 2004.



%\bibitem{Chen:CS}
%C.~-Y.~Chen, C.~-H.~Wang, and C.~-C.~Chao,
%``Complementary sets and Reed-Muller codes for peak-to-average power ratio reduction in OFDM,''
%\emph{in Proc. of 16th AAECC, Lecture Notes in Computer Science},
%vol.~3857, pp.~317-327, 2006.








%\bibitem{Elia:rest}
%S.~Eliahou, M.~Kervaire, and B.~Saffari,
%``A new restriction on the lengths of Goay complementary sequences,''
%\emph{J. Combin. Theory Ser. A}, vol.~55, pp.~49-59, 1990.




%\bibitem{Fied:frame}
%F.~Fiedler, J.~Jedwab, and M.~G.~Parker,
%``A framework for the construction of Golay sequences,''
%\emph{IEEE Trans. Inform. Theory}, vol.~54, no.~7, pp.~3113-3129, July 2008.









%\bibitem{MassMitt:Welch}
%J.~L.~Massey and T.~Mittelholzer,
%``Welch's bound and sequence sets for code-division multiple-access systems,''
%in \emph{Sequences II: Methods in Communication, Security, and Computer Science,}
%R.~Capocelli, A.~De Santis, and U.~Vaccaro (eds.),
%pp.~63-78, New York: Springer-Verlag, 1993.







%\bibitem{ParkTell:bin}
%M.~G.~Parker and C.~Tellambura,
%``A construction for binary sequence sets with low peak-to-average power ratio,''
%\emph{Report in Informatics}, no.~242, Department of Informatics, University of Bergen, Bergen, Norway, Feb. 2003.






%\bibitem{Pop:syn}
%B.~M.~Popovi\'c,
%``Synthesis of power efficient multitone signals with flat amplitude spectrum,''
%\emph{IEEE Trans. Commun.}, vol.~39, pp.~1032-1033, 1991.




%\bibitem{Rud}
%W.~Rudin,
%``Some theorems on Fourier coefficients,''
%\emph{Proc. Amer. Math. Soc.,} no.~10, pp.~855-859, 1959.







%\bibitem{Sarw:MWBE}
%D.~V.~Sarwate,
%``Meeting the Welch bound with equality,''
%\emph{Sequences and Their Applications: Proceedings of SETA'98},
%C.~Ding, T.~Helleseth, and H.~Niederreiter (eds.),
%DMTCS series, Springer-Verlag, 1999.




%\bibitem{Schmidt:coset}
%K.~-U.~Schmidt,
%``On cosets of the generalized first-order Reed-Muller code with low PMEPR,''
%\emph{IEEE Trans. Inform. Theory}, vol.~52, no.~7, pp.~3220-3232, July 2006.

%\bibitem{Schmidt:gen}
%K.~-U.~Schmidt,
%``Complementary Sets, Generalized Reed-Muller Codes, and Power Control for OFDM,''
%\emph{IEEE Trans. Inform. Theory}, vol.~53, no.~2, pp.~808-814, Feb. 2007.




%\bibitem{Shap}
%H.~S.~Shapiro,
%``Extremal problems for polynomials,``
%\emph{M.S.~Thesis}, MIT, 1951.





%\bibitem{Turyn:Had}
%R.~J.~Turyn,
%``Hadamard matrices, Baumert-Hall units, four-symbol sequences, pulse compression,
%and surface wave encodings,''
%\emph{J.~Combin.~Theory (A)}, vol.~16, pp.~313-333, 1974.


%\bibitem{Welch:lower}
%L.~R.~Welch,
%``Lower bounds on the maximum cross correlation of signals,''
%\emph{IEEE Trans. Inform. Theory}, vol.~IT-20, no.~3, pp.~397-399, May 1974.














%\bibitem{McE:coding}
%R.~J.~McEliece,
%\emph{The Theory of Information and Coding},
%Encyclopedia of Mathematics and Its Applications, vol.~3,
%Addison-Wesley, Reading, Massachustts, 1977.


\end{thebibliography}

\end{document}